\pdfoutput=1
\documentclass{article}
\PassOptionsToPackage{numbers, compress}{natbib}

\usepackage[preprint, eandd]{neurips_2026}

\usepackage[utf8]{inputenc}
\usepackage[T1]{fontenc}
\usepackage{hyperref}
\usepackage{url}
\usepackage{booktabs}
\usepackage{amsfonts}
\usepackage{amssymb}
\usepackage{nicefrac}
\usepackage{microtype}
\usepackage{xcolor}
\usepackage{graphicx}
\usepackage{placeins}
\usepackage{float}
\usepackage{enumitem}
\usepackage[singlelinecheck=false]{caption}

\title{A Stratified Multi-Rater Evaluation of LLM-Based Virtual Standardized Patients\\
with a Deployed Data-Generation Platform}

\author{%
  Yongjin Yi\thanks{Equal contribution.} \\
  Internal Medicine, College of Medicine, \\
  Dankook University \\
  \texttt{yongjin.neph@gmail.com} \\
  \And
  Jinyong Park\footnotemark[1] \\
  Department of Medical Device Development, \\
  Seoul National University \\
  \texttt{nwirandx@gmail.com} \\
  \And
  Sejoong Kim \\
  Internal Medicine, College of Medicine, \\
  Seoul National University \\
  \texttt{sejoong2@snu.ac.kr} \\
}

\begin{document}

\maketitle

\begin{abstract}
The Objective Structured Clinical Examination (OSCE) is widely used to assess clinical communication skills in medical education, requiring trainees to take histories and reason clinically with standardized patients (SPs). Large language model (LLM)-based virtual standardized patients (VSPs) offer a practical alternative to human SPs. These are typically built using persona-injected prompting, which casts a model into a clinical role via a single system prompt. However, existing evaluations of such VSPs often rely on a single backbone LLM, small human rater panels, or LLM-as-a-judge scoring without an utterance-aligned human-expert reference. Consequently, they are limited in their ability to identify the specific shortcomings of individual models. In this study, we evaluated four LLMs (GPT-4o, Claude 3.5 Sonnet, Llama-3.1 70B, and HyperCLOVA X), spanning commercial, open-source, and Korean-developed models, against a human-expert reference across three Korean OSCE cases (hematuria, fever in pregnancy, and chronic cough). In a double-blind, fully crossed design, 17 senior medical-student raters scored every model utterance on three sentence-level and six encounter-level Likert metrics, yielding 7{,}480 data points. We analyzed the scores using a stratified linear mixed-effects model based on an 11-category history-taking and 3-level linguistic-form taxonomy. At the encounter level, three of the four LLMs fell 0.08 to 0.47 Likert points below the human-expert reference on Comprehensibility, Consistency and Hallucination-freeness. The difference was not statistically significant for Hallucination-freeness in Claude 3.5 Sonnet and GPT-4o, Consistency in GPT-4o, and Engagement in HyperCLOVA X. Llama-3.1 70B fell significantly below the reference on all six metrics. Furthermore, we introduce a web-based OSCE platform that supports virtual practice across diverse clinical cases. This platform continuously collects student--VSP dialogues under research and privacy consent; the resulting dataset is available at \url{https://huggingface.co/datasets/nwirandx/Korean-OSCE-history-taking-dialogues}.

\end{abstract}

\section{Introduction}
The Objective Structured Clinical Examination (OSCE) has been the gold
standard for assessing clinical skills (history taking, communication,
clinical reasoning, and patient education) in medical education for
decades [1]. National licensure programs in Taiwan, Canada, Germany,
Spain, and the Republic of Korea have incorporated OSCE-format clinical
examinations [2, 3], and OSCE-style training has been shown to shift
student preparation toward integrated clinical competencies [4].
Despite this established role, the OSCE administration imposes
substantial costs and accessibility barriers. Standardized patient (SP)
programs require sustained investment in recruitment, training, examiner
panels, and dedicated facilities. These costs, together with concerns
about discriminative validity given consistently high pass rates,
contributed to the 2021 retirement of the United States Medical
Licensing Examination (USMLE) Step 2 Clinical Skills exam [5]. In the
same year, the Medical Council of Canada terminated the MCCQE Part II
OSCE for similar reasons of cost-effectiveness and redundancy with
residency-based competency assessments [6, 7]. Although the
educational rationale for the OSCE remains intact [8], the
discontinuation of national clinical-skills examinations has
necessitated cost-effective alternatives that preserve the formative
function of OSCE-based training.

Large language models (LLMs) have rapidly emerged as a candidate
substrate for such alternatives. Frontier models encode sufficient
clinical knowledge to surpass the USMLE passing threshold [9] and
approach expert performance on USMLE-style soft-skills items covering
ethics, communication, and empathy [10]. Building on this capacity,
several groups have implemented LLM-based virtual standardized patients
(VSPs) via \emph{persona injection}, a system-prompt-based approach in which
the model is instructed to play a clinical role and answer utterance by
utterance from within it. Despite its technical simplicity, this
approach is plausible enough to recover several pedagogically important
behaviors of human SPs, opening the prospect of deployable,
low-marginal-cost training environments for clinical communication.

Research on LLM-based VSPs has progressed along three axes: feasibility
and usability in history-taking practice [11, 12], educational
efficacy demonstrated through randomized controlled trials of
LLM-simulated encounters [13, 14], and agent-level dialogic
capabilities approaching those of primary-care physicians [15]. Two
recent syntheses [16, 17] conclude that the field is maturing rapidly,
but that current evidence is heterogeneous in evaluation methodology and
concentrated in English-language, single-scenario studies. Despite this growth, several gaps remain in the evaluation of LLM-based
VSPs. Most studies have evaluated a single LLM on a single scenario,
often with small rater panels and without formal inter-rater reliability
analysis, so fully-crossed multi-rater comparisons against a
human-expert baseline on identical clinical tasks are scarce. Existing
reports also rely largely on aggregate Likert means or single composite
scores, which obscure where a model fails: a single rank cannot
distinguish a model that mishandles greetings from one that
misinterprets clarifying questions about symptom character. In addition,
almost all published LLM-VSP evaluations have been conducted in English,
although diagnostic dialogue is constructed across many languages and
clinical traditions; whether the same persona-injection prompt yields
comparable behavior when the question is phrased in a language whose
yes/no semantics differ from English is largely untested. The deployable
utility of LLM-based VSPs relative to human SPs on specific dimensions
of clinical communication remains an open question.

We address these gaps with a stratified, multi-rater human-evaluation
framework for LLM-based VSPs, accompanied by a deployed training
platform that enables the same evaluation to extend beyond a static set
of prompts. Our work makes the following contributions.

\begin{enumerate}[leftmargin=2em,itemsep=2pt,topsep=4pt,parsep=0pt]
\item We report a fully-crossed multi-rater evaluation against a human-expert reference. We evaluate four frontier LLMs (GPT-4o, Claude 3.5 Sonnet, Llama-3.1 70B, HyperCLOVA X) alongside a human medical-education expert serving as the SP-equivalent reference, across three OSCE cases. Seventeen final-year medical-student raters produced 7,480 sentence- and encounter-level ratings under formal inter-rater reliability analysis.

\item Every medical student input prompt is tagged with an 11-category history-taking taxonomy grounded in AMEE Guide No. 81 [1] and with a three-level linguistic-form label (affirmative, negative, open-ended). The taxonomy surfaces failure modes that aggregate Likert scores would otherwise hide, including a Korean negative-form penalty on Fluency.

\item We deploy a real-time persona-injected VSP web application on which medical students conduct free-form OSCE-style encounters, and we show that the same taxonomy applies to dialogue collected on the platform. The platform-collected corpus is released under informed consent on a public Hugging Face repository with Croissant Core and Responsible-AI metadata.
\end{enumerate}

\section{Related Work}
\paragraph{LLM-based virtual standardized patients.}

LLM-based virtual standardized patients have most often been built using a single declarative system prompt, called persona injection, that fixes the patient's identity, clinical state, history, and conversational style, and constrains turn-by-turn answers to remain in role [18]. Within this paradigm, several representative systems anchor the field. Holderried et al. [11] developed a GPT-powered simulator for abdominal pain history-taking and validated it through mixed-methods qualitative interviews with German medical students. Elhilali et al. [12] developed a communication-skills simulator with user-centered usability as the principal outcome. Br\"ugge et al. [13] conducted a randomized controlled trial in which structured feedback following LLM-simulated encounters improved medical students' clinical decision-making, as assessed against expert rubrics. Lavigne et al. [14] reported the first deployment-scale RCT (n = 247) of an LLM-based OSCE-training platform (DocSimulator) in \emph{NEJM AI}, using faculty-graded OSCE performance as the primary endpoint. At the agent-capability frontier, Tu et al. [15] introduced AMIE, a self-play--trained diagnostic-dialogue agent that matched or exceeded primary-care physicians across history taking, diagnostic reasoning, management, and patient-perceived empathy in standardized patient-actor encounters. Most recently, Kyung et al. [19] introduced PatientSim, a persona-driven simulator comprising 37 personas, validated by four clinicians. Existing LLM-VSP studies typically score model output against rubric items or encounter-level patient-actor impressions, evaluate a single backbone LLM, and rely on small or single-rater panels. The present evaluation, in contrast, compares every model utterance against an utterance-aligned response from a practicing medical educator as a 1:1 gold standard, scores every utterance under a fully-crossed assignment by seventeen OSCE-trained final-year medical-student raters, and stratifies the resulting scores along a history-taking and linguistic-form taxonomy that localizes where each model incurs Likert-scale penalties. To our knowledge, this methodological combination has not been instantiated in prior LLM-VSP research.

\paragraph{Multi-rater human evaluation of clinical dialogue agents.}

Educational simulators differ from closed-form benchmarks in that there is no single correct answer to score against; quality is multi-dimensional, partly subjective, and context-bound. Aggregate accuracy or single-judge LLM-as-a-judge scores, therefore, underdetermine the educational utility of a VSP. Multi-rater human evaluation has become the gold standard whenever judgments are inherently subjective: the DICES dataset [20] uses 123 raters per dialogue to model rater diversity rather than treat disagreement as noise, and HealthBench [21] engages 262 physicians from 60 countries to score 5,000 multi-turn health conversations.

\paragraph{Stratified, capability-based evaluation.}

The lineage initiated by CheckList [22] and broadened by HELM [23] argues that systems should be assessed not only on overall scores but on category-defined failure modes, and that publishing diagnostic detail is itself a contribution. Within medicine, Yao et al. [24] introduced MedQA-CS as an OSCE-style closed-form benchmark with Cohen's $\alpha$ agreement, and Jiang et al. [25] built MedAgentBench as an EHR-agent environment validated by two physicians. CLEVER [26] provides a methodology for blind preference-based evaluation by external clinicians. We combine 1) fully-crossed multi-rater scoring of LLM-VSP dialogue, 2) explicit history-taking and linguistic stratification of the question itself, and 3) analysis of how stratification surfaces model-specific failure modes that aggregate scores hide. The framework also transfers from static evaluation to dialogue collected on a deployed platform.

\section{VSP Construction and Response Generation}
A virtual standardized patient evaluated against an OSCE encounter must
satisfy three requirements: it must inhabit a clinically coherent
persona consistent with a specific scenario, respond to free-form
student utterances in role and in real time, and do so reproducibly so
that responses are comparable across systems. We construct VSPs that
meet these requirements by composing OSCE scenarios written by a
clinical educator, a single declarative persona-injection prompt that
instantiates each scenario, and a multi-source response corpus in which
four frontier LLMs and a human medical-education expert independently
produce responses to the same input utterances.

\subsection{OSCE Scenarios and Input Prompts}

Three OSCE scenarios were developed: hematuria (red urine), fever in a
pregnant patient, and chronic cough. Each was developed from the Korean
OSCE/CPX training guide for medical students [27] and refined by a board-certified internal
medicine faculty member with thirteen years of experience in clinical
skills assessment. The cases span a common general-medicine presentation
(hematuria), a special-population case requiring sensitive
history-taking (pregnancy), and a chronic presentation requiring
multi-specialty differential reasoning (cough). Each scenario specifies
the patient's chief complaint, demographic profile, history of present
illness, past medical, drug, family, and social history, and the
expected behavioral style of the SP (e.g., reluctance to disclose
particular history). Scenarios are stored as structured documents that map directly to the persona-injection prompt fields used in the next subsection.

A study-specific set of seventy input prompts (23 hematuria, 23 fever,
24 cough) was prepared as the queries against which all VSPs are
evaluated. These prompts are student-doctor utterances sampled from
authentic OSCE-style training transcripts produced in prior cohorts at
the same medical school and represent the question types that medical
students typically deploy during a 10-minute OSCE encounter, namely
opening greetings, chief-complaint elicitation, character and
associated-symptom probing, past-medical and lifestyle inquiry,
summarization, and patient education. Each prompt is independently
tagged with a history-taking category and a linguistic-form label, as
detailed in \S4.3. All scenarios and the input-prompt set were reviewed
and approved by the Institutional Review Board of Dankook University
Hospital prior to data collection.

\subsection{VSP Prompting via Persona Injection}

Each VSP is constructed by \emph{persona injection} [18], a
system-prompt-based approach in which a single declarative prompt
instructs the model to play a specific clinical role and to answer
turn-by-turn from within it. The same template, comprising six fields
(role declaration, demographic context, clinical state, history,
conversational style, and behavioral constraints; full template in
Appendix B), was instantiated three times, once per scenario, with
case-specific values; no other prompt text varied between scenarios.
Each instantiation is on the order of 700--900 Korean characters and is
released verbatim with the dataset.

Inference parameters were held constant across the four LLMs:
temperature 0.7, top-\emph{p} 0.95, maximum response length 1,024 tokens, and
a single response per query (no retries, no chain-of-thought
scaffolding). The four LLMs evaluated were GPT-4o (OpenAI), Claude 3.5
Sonnet (Anthropic), Llama-3.1 70B Instruct (Meta, served via a
HuggingFace Inference Endpoint), and HyperCLOVA X (NAVER, served via the
official Korean-language API). Specific dated API releases are omitted;
all four models were queried within a single calendar week to minimize
drift across silent updates.

\subsection{Generating the Multi-Source Response Corpus}

For each of the seventy input prompts, five sources independently
produced a response: the four LLMs above and one human medical-education
expert (the same internal-medicine faculty member who authored the scenarios). Expert responses were composed without exposure to
model outputs (write-blind condition) and at the level of detail
expected from a trained standardized patient at a national OSCE
workshop. The expert response set serves as a 1:1 utterance-aligned gold
standard against which model outputs are directly compared.

The complete multi-source response corpus consists of 5 sources $\times$
70 input prompts = 350 responses. Responses were stored together with
their source label, randomized within case for rater presentation to
mitigate position-order bias, and re-blinded such that raters did not
know which response came from which source.

\section{Stratified Multi-Rater Evaluation}
A multi-rater human evaluation framework was applied to the 350-response
corpus described above. Seventeen raters scored every response on three
sentence-level metrics (Fluency, Accuracy, Contextual Relevance) and
every encounter on six encounter-level metrics. The fully-crossed design
produced 5,950 sentence-level rating events (17 raters $\times$ 350
responses) and 1,530 encounter-metric assessments (17 raters $\times$ 15
encounters $\times$ 6 metrics), totalling 7,480 distinct rating actions.
Resulting scores were stratified along an 11-category history-taking
taxonomy crossed with a 3-level linguistic-form label and analyzed using
inter-rater reliability statistics and a linear mixed-effects model.

\subsection{Rater Recruitment}

Seventeen senior medical students were recruited from Korean medical
schools via an open invitation circulated through the medical-education
office (mean age 24.8 $\pm$ 1.9 years; 5 women [29.4\%]; Year 6 n = 10
[58.9\%], Year 5 n = 5 [29.4\%], Year 4 n = 2 [11.8\%] of the 6-year
curriculum). All had completed at least two prior OSCE training cycles,
ensuring familiarity both with the OSCE format and with the educational
target of an SP-style encounter.

A 60-minute calibration session preceded the rating phase, in which the
rubric (\S4.2) was reviewed against three illustrative responses (one
human-expert, one strong-LLM, one weak-LLM) drawn from a held-out
scenario; raters discussed disagreements until anchor descriptions for
each Likert level were shared. Raters then scored the full corpus
asynchronously over a four-week window in randomized order and with
sources de-identified.

\subsection{Sentence- and Encounter-Level Metrics}

Two granularities of evaluation are used (Table 1). At the sentence
level (one rating per response, where a response corresponds to one VSP
utterance), each rater scored Fluency, Accuracy, and Contextual
Relevance on five-point Likert scales anchored to written descriptors at
each level. At the encounter level (one rating per scenario $\times$
source pair), six metrics were used to capture properties that emerge
only from the full conversational arc: comprehensibility,
hallucination-free, coherence, consistency, engagement, and overall
satisfaction. The complete rubric is reproduced in Appendix C.

\begin{table}[!htbp]
  \centering\footnotesize
  \caption{Sentence- and encounter-level evaluation metrics. The ``Rater item'' column shows the exact Likert agreement statement presented to raters (translated from Korean), rated on a 5-point scale (1 = strongly disagree; 5 = strongly agree).}
  \label{tab:metrics}
  \begin{tabular}{p{0.3\linewidth}p{0.6\linewidth}}
    \toprule
    \textbf{Metric} & \textbf{Rater item} \\
    \midrule
    \multicolumn{2}{l}{\textit{Sentence-level (per VSP utterance)}} \\
    \quad Fluency & The response reads naturally, as if produced by a human. \\
    \quad Accuracy & The response is factually accurate based on the provided patient case description. \\
    \quad Contextual Relevance & In the OSCE setting, the response accurately addresses the intent of the student's question. \\
    \midrule
    \multicolumn{2}{l}{\textit{Encounter-level (per scenario $\times$ source pair)}} \\
    \quad Comprehensibility & The virtual patient's responses are clear and easy to understand. \\
    \quad Hallucination-free & The virtual patient does not provide content not given in the case description. \\
    \quad Coherence & The dialogue flow is natural and coherent across turns. \\
    \quad Consistency & The persona is consistent across the encounter. \\
    \quad Engagement & The conversation feels engaging, comparable with a human SP. \\
    \quad Overall Satisfaction & Overall, the encounter would be acceptable as a substitute for a human SP. \\
    \bottomrule
  \end{tabular}
\end{table}

\subsection{Question Stratification: History-Taking Function and Linguistic Form}

We adopt the stratified-evaluation paradigm, in which models are
assessed on category-defined slices of input rather than aggregate
scores alone, as introduced by CheckList [22] and broadened by HELM
[23]. Each input prompt is annotated along two independent axes,
which, when crossed, reveal stratified contrasts that aggregate scores
cannot detect. The first axis is an 11-category history-taking taxonomy
grounded in the OSCE history-taking structure of AMEE Guide No. 81 [1]
and validated against the systematic review of medical-student
history-taking instruction by Keifenheim et al. [28] (Table 2). The
second is a three-level linguistic-form label distinguishing
\emph{Affirmative} (positive-polarity questions and statements), \emph{Negative}
(negative-polarity questions), and \emph{Open-ended} (unconstrained-response
questions). The Affirmative--Negative distinction is motivated by the
polarity-response divergence between Korean and English, which is the failure
mode that the linguistic axis is designed to surface. Both axes were
applied with a rule-based tagger; ambiguous utterances were reviewed by
two faculty members and resolved by consensus. The two-axis full cross
yields an 11 $\times$ 3 stratification grid that supports cell-level
aggregation, marginal analysis, and interaction modeling.

\begin{table}[!htbp]
  \centering\footnotesize
  \caption{Eleven-category history-taking taxonomy of input prompts.}
  \label{tab:typology}
  \begin{tabular}{p{0.20\linewidth}c p{0.30\linewidth} p{0.34\linewidth}}
    \toprule
    \textbf{Category} & \textbf{\# Q} & \textbf{Definition} & \textbf{Example} \\
    \midrule
    Greeting \& ID check & 3 & Opening of the encounter and verification of patient identity. & \emph{``Hello, I'm a medical student. May I have your name and age?'' (hematuria)} \\
    Chief complaint (C.C.) & 3 & Eliciting the principal reason for the visit. & \emph{``Hello, what brings you in today?'' (hematuria)} \\
    History of C.C.\newline (HPC) --- Character & 10 & Probing the qualitative character of the present complaint. & \emph{``What color was the red urine?'' (hematuria)} \\
    HPC ---\newline Associated symptoms & 21 & Eliciting symptoms accompanying the chief complaint. & \emph{``Phlegm or runny nose along with the cough?'' (cough)} \\
    HPC ---\newline Aggravating/relieving & 3 & Contextual factors that worsen or improve the symptom. & \emph{``Does the cough get worse when you lie down?'' (cough)} \\
    Past medical history & 6 & Previous diagnoses and prior episodes. & \emph{``Diagnosed with diabetes before?'' (fever)} \\
    Drug history & 3 & Current and recent medications. & \emph{``Taking any medications regularly?'' (fever)} \\
    Family history & 2 & Hereditary conditions relevant to differential. & \emph{``Anyone in family with lung cancer?'' (cough)} \\
    Social history & 9 & Tobacco, alcohol, occupation, lifestyle. & \emph{``Do you drink alcohol or smoke?'' (cough)} \\
    Clinical reasoning & 2 & Summarization and check-back. & \emph{``To summarize: cold one week ago, then reddish-brown urine\dots'' (hematuria)} \\
    Patient education & 8 & Diagnosis explanation, counseling, lifestyle guidance. & \emph{``Hematuria may be associated with smoking; I recommend quitting.'' (hematuria)} \\
    \bottomrule
  \end{tabular}
\end{table}

\subsection{Inter-Rater Reliability and Mixed-Effects Modeling}

\textbf{Inter-rater reliability} was quantified at both sentence and encounter levels by ICC(2,k) (two-way random-effects, absolute agreement, average measures; following Koo and Li [29]) and by Krippendorff's $\alpha$ with ordinal weights [30]; ICC values are reported with 95\% confidence intervals (F-distribution method).

\textbf{Statistical comparison of model performance} was carried out by a
linear mixed-effects model [31] fitted to rater-level Likert scores.
The model included \emph{Source} (5 levels), \emph{History-taking category} (11
levels), and \emph{Linguistic form} (3 levels) as fixed effects, with
\emph{Source} $\times$ \emph{History-taking category} and \emph{Source} $\times$
\emph{Linguistic form} interactions to expose stratified contrasts; \emph{Rater}
was modeled as a random intercept and \emph{Utterance} as a random intercept
nested within \emph{Case}. Estimation was by restricted maximum likelihood
(REML); estimated marginal means and contrasts between Source levels
(and across the \emph{a priori} Negative-vs-Affirmative contrast within each
Source) are reported with 95\% confidence intervals. Descriptive means
reported in Table~3, Appendix D, and Figure E1 are accompanied by 95\%
nonparametric percentile bootstrap confidence intervals (2{,}000
resamples). All analyses were performed in
R 4.5.2 and Python 3.11.

\textbf{Data availability.} All static-evaluation materials, namely the
350-response corpus, the 7,480 rater scores, the input-prompt set with
two-axis annotations, and the persona-injection prompt template, are
provided in full in the Appendices (Appendices B-C). The static
evaluation set is not separately hosted as a stand-alone public dataset
given its modest size and the human-expert-reference component. The
platform-collected dialogue corpus is released separately as a
public Hugging Face dataset under CC BY-NC-SA 4.0 with a Croissant Core
and Responsible-AI metadata file at \url{https://huggingface.co/datasets/nwirandx/Korean-OSCE-history-taking-dialogues}. The Datasheet for Datasets (Appendix A) follows Gebru et al. [32] and documents the platform-collected dataset.

\section{Experimental Results}
\subsection{Encounter-level performance}

The human-expert reference led on all six encounter-level metrics (Table
3). Mixed-effects contrasts against the human reference (Figure 1)
showed that three of the four LLMs (Claude 3.5 Sonnet, GPT-4o,
HyperCLOVA X) stayed within 0.5 Likert
points of the reference on Comprehensibility, Consistency, and
Hallucination-freeness. In four of the 24 source $\times$ metric cells
the contrast was not statistically significant: Hallucination-freeness
in Claude 3.5 Sonnet and GPT-4o, Consistency in GPT-4o, and Engagement
in HyperCLOVA X (-0.02 [-0.31, 0.27]). Llama-3.1 70B fell significantly
below the reference on all six metrics and trailed every other source,
with the largest gap on Coherence (-1.86 [-2.14, -1.59])
and Overall Satisfaction showing an additional 1.5--2 point spread below
the reference.

\begin{table}[!htbp]
  \centering\scriptsize
  \setlength{\tabcolsep}{3.6pt}
  \caption{Encounter-level mean Likert score [95\% bootstrap CI] for each source, pooled across the three OSCE cases. Each cell pools 51 ratings (17 raters $\times$ 3 cases).}
  \label{tab:enc}
  \begin{tabular}{lcccccc}
    \toprule
    \textbf{Source} & \textbf{Compre-} & \textbf{Hallucination-} & \textbf{Coherence} & \textbf{Consistency} & \textbf{Engagement} & \textbf{Overall} \\
                    & \textbf{hensibility} & \textbf{free} & & & & \textbf{satisfaction} \\
    \midrule
    \textbf{Human (reference)} & \textbf{4.84 [4.74--4.94]} & \textbf{4.49 [4.29--4.67]} & \textbf{4.80 [4.69--4.92]} & \textbf{4.75 [4.57--4.88]} & \textbf{4.18 [3.92--4.43]} & \textbf{4.61 [4.41--4.78]} \\
    Claude 3.5 Sonnet & 4.45 [4.27--4.63] & 4.35 [4.12--4.55] & 4.00 [3.75--4.25] & 4.39 [4.16--4.61] & 3.82 [3.55--4.10] & 3.82 [3.53--4.12] \\
    GPT-4o & 4.59 [4.39--4.76] & 4.41 [4.22--4.61] & 4.08 [3.84--4.29] & 4.53 [4.35--4.71] & 3.75 [3.49--3.98] & 3.82 [3.57--4.08] \\
    Llama-3.1 70B & 3.71 [3.43--3.98] & 3.94 [3.69--4.18] & 2.94 [2.67--3.24] & 3.82 [3.53--4.14] & 2.94 [2.63--3.25] & 2.71 [2.35--3.06] \\
    HyperCLOVA X & 4.57 [4.41--4.71] & 4.02 [3.78--4.25] & 4.24 [4.04--4.43] & 4.39 [4.24--4.55] & 4.16 [3.92--4.37] & 3.88 [3.65--4.10] \\
    \bottomrule
  \end{tabular}
\end{table}

\begin{figure}[!htbp]
  \centering
  \includegraphics[width=\linewidth]{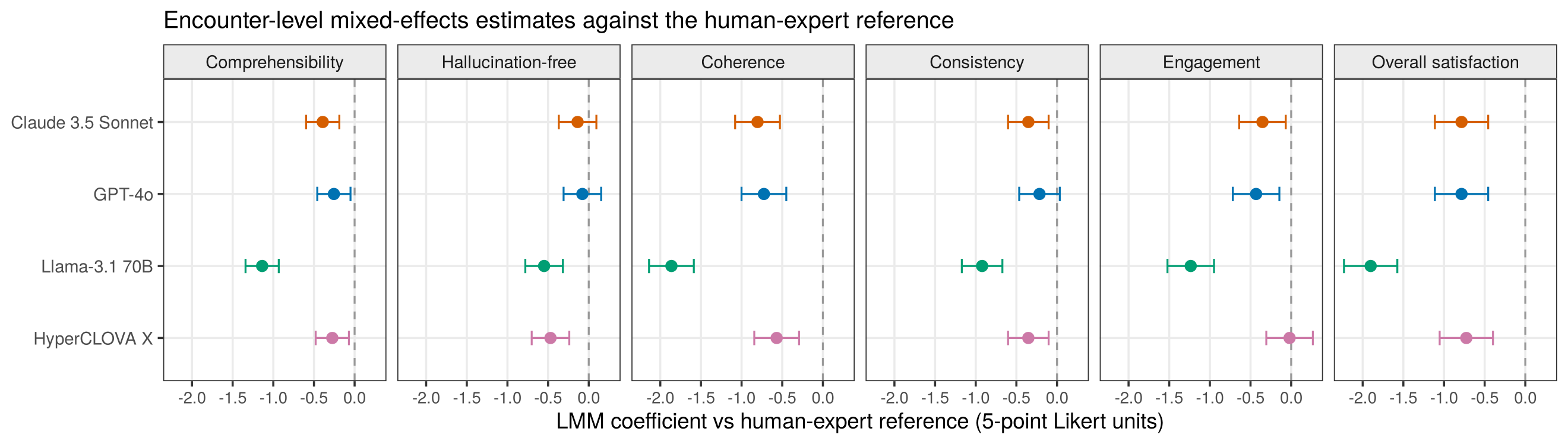}
  \caption{Linear mixed-effects model estimates (mean difference vs.\ human-expert reference, in 5-Likert units) with 95\% confidence intervals for each LLM $\times$ encounter-level metric. Negative values indicate lower than the human reference; the reference (Human) sits at 0 by construction.}
  \label{fig:forest}
\end{figure}

\subsection{Sentence-level performance}

Sentence-level performance was substantially higher and more compressed
across sources than encounter-level (Figure 2). Pooled Accuracy was
within 0.09 Likert points of the human reference across all four LLMs,
indicating that within-utterance factual alignment is the most
achievable property. Fluency separated the sources most clearly: human
reference 4.91, HyperCLOVA X 4.68, GPT-4o 4.39, Claude 3.5 Sonnet 4.31,
Llama-3.1 70B 3.69 of 5. The Fluency ordering anticipates the
encounter-level Coherence ordering (Figure 1), suggesting that local
linguistic naturalness propagates into global encounter quality.

\begin{figure}[!htbp]
  \centering
  \includegraphics[width=\linewidth]{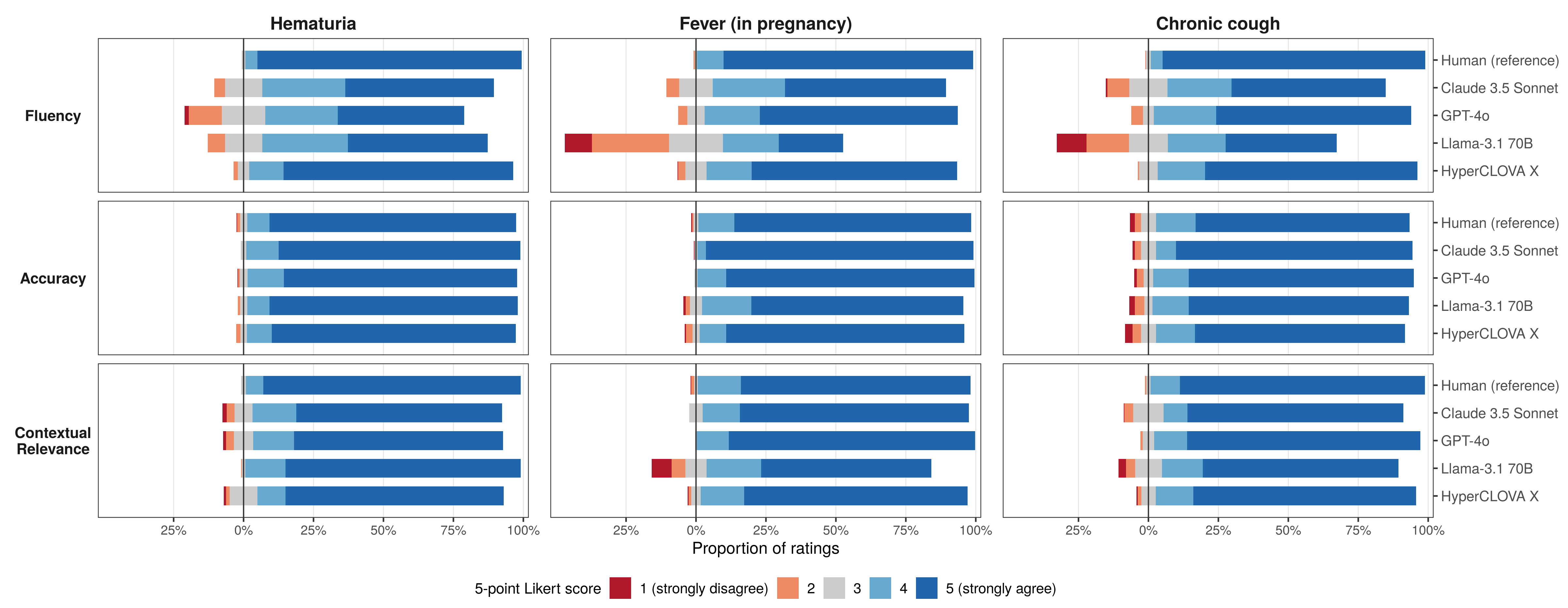}
  \caption{Diverging stacked-bar distribution of Likert ratings for the three sentence-level metrics for each of the five sources, shown for each of the three OSCE cases. The vertical zero line splits ratings into negative (1--2 plus half of 3) on the left and positive (4--5 plus half of 3) on the right.}
  \label{fig:likert}
\end{figure}

\subsection{Stratified analysis by question typology}

Question-typology stratification revealed model- and category-specific
failure modes that aggregate scores obscured. History-taking category
$\times$ Source interactions (Figure 3) identified pronounced LLM
weaknesses on Clinical reasoning and Patient education, categories that
require synthesis across the prior dialogue rather than direct symptom
recall. They also revealed that Llama-3.1 70B's Fluency deficit exceeded
1 Likert point below the human reference in seven of the eleven
categories, with the largest gaps on Chief complaint and Patient
education. Linguistic-form $\times$ Source
interactions surfaced a Korean negative-form penalty on Fluency.
Relative to the human reference, the additional drop from
Affirmative-form questions to negative-polarity yes/no questions was
0.149 [-0.286, -0.011] for Llama-3.1 70B, 0.124 [-0.261, 0.014] for
GPT-4o, 0.093 [-0.231, 0.044] for HyperCLOVA X, and 0.033 [-0.170,
0.105] for Claude 3.5 Sonnet; the human reference itself showed no such
directional effect (+0.043 [-0.067, 0.153]). Among the four models,
only the Llama-3.1 70B interval excludes zero; the remaining three are
directional but not statistically significant. The ordering does not track
Korean-language ability: the penalty is smallest for Claude 3.5 Sonnet and
larger for HyperCLOVA X, the only Korean-developed model. The effect is one
of question form, not of proficiency. Per-form
means by source are visualized in Appendix E (Figure E1).

Sixteen of the 350 utterances fell below 3.0 on at least one
sentence-level metric, and fourteen of these were Fluency failures. In
seven utterances, all in Clinical reasoning or Patient education, the
virtual patient restated the student's explanation in clinical register
instead of answering as a patient. In three, all negative-polarity
questions, it answered a question that had not been asked. No
human-reference utterance fell below 3.0 on any metric. These patterns
thus distinguish the model-generated responses from the human reference.

\begin{figure}[!htbp]
  \centering
  \includegraphics[width=\linewidth]{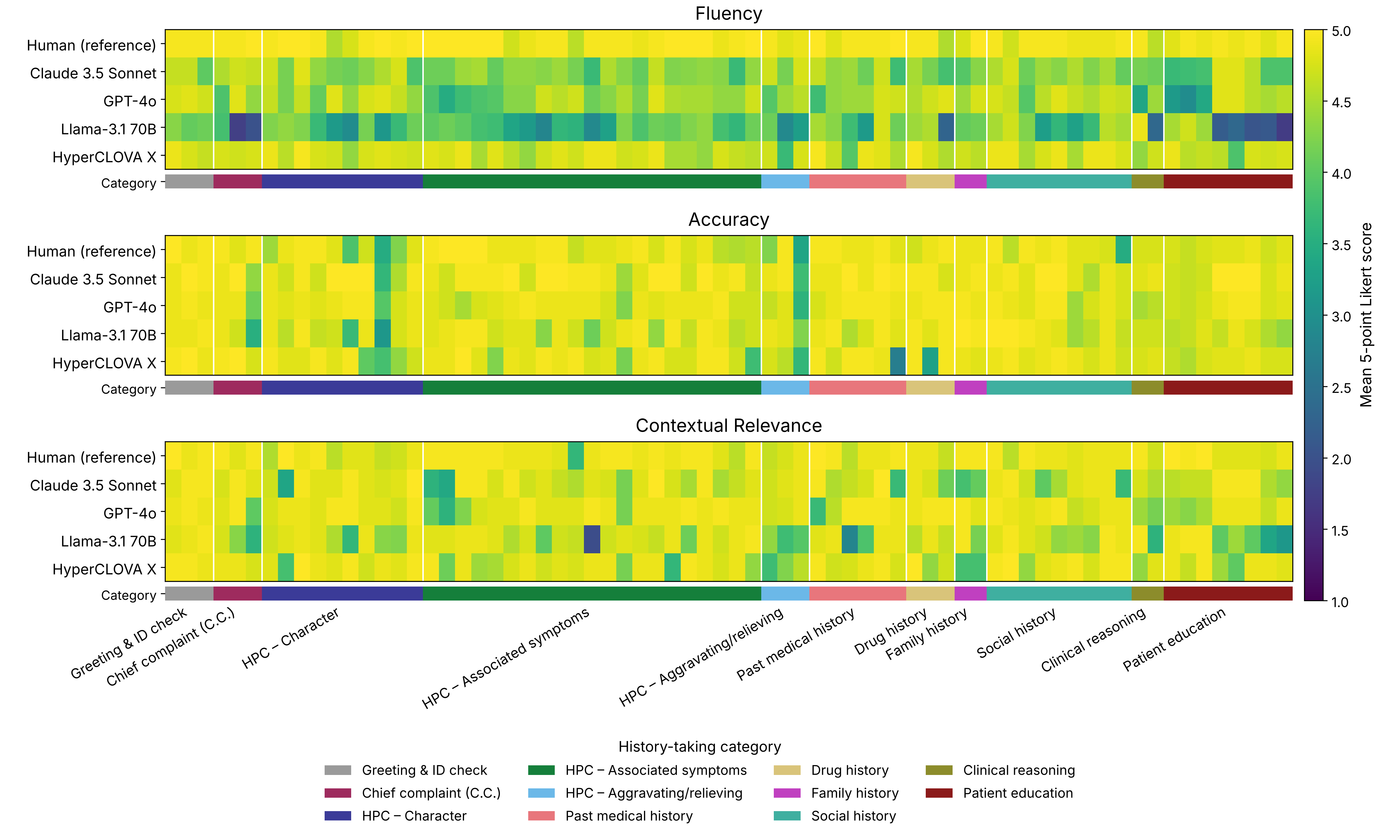}
  \caption{Sentence-level mean Likert score per question $\times$ source, organized by sentence-level metric (Fluency, Accuracy, Contextual Relevance). Within each metric panel, rows are the five sources (Human reference + four LLMs) and columns are the 70 input prompts sorted by their history-taking category; cell shading shows mean Likert score (1--5). The horizontal color bar (\textit{Category}) below each metric panel maps each prompt to its history-taking category (legend at bottom).}
  \label{fig:heatmap}
\end{figure}

\subsection{Inter-rater reliability}

Encounter-level pooled ICC(2,k) ranged from 0.79 (Hallucination-free) to 0.94 (Coherence), with a fully-pooled ICC(2,k) of 0.91 [0.88--0.94]. Sentence-level pooled ICC(2,k) per source ranged from 0.80 to
0.92. Krippendorff's $\alpha$ with ordinal weights was lower than ICC at
both granularities (0.08--0.41), reflecting $\alpha$'s stricter
accounting for chance agreement on ordinal scales; the two measures
produced the same source-level ordering. Full per-metric and per-source
values are reported in Appendix E (Table E1).

\section{VSP-OSCE Training Platform}
\subsection{Platform architecture and scenario pipeline}

The \textbf{VSP-OSCE training platform} is a web-based application that
delivers persona-injected VSPs as a continuously updating educational
resource and dataset pipeline. The platform invokes GPT-4.1 for
clinical-scenario script generation (offline, prior to deployment) and
GPT-4o for the two real-time roles, namely automated evaluation and
patient role-play (chosen for its lower latency during live student
sessions); all calls use seed = 42. Scenario content is
derived from a structured database of 80 core symptoms commonly taught
in medical curricula, linked to expert-validated disease lists and three
patient profiles per symptom (240 profiles in total) covering
demographics, vital signs, and diagnostic context, organized in
hierarchical JSON, with symptoms at the top, diseases nested under each
symptom, and patient profiles nested under each disease. For each
disease, a symptom-focused history-taking checklist is generated by
GPT-4.1 (temperature 0.2) under a one-shot prompt and refined through
human-in-the-loop expert review; a free-form patient narrative ($\geq$
1,500 characters) is then generated and verified by a downstream
diagnostic model that retains the scenario only when its predicted
disease matches the intended diagnosis. At session start, the verified scenario is loaded into the persona-injection prompt template (Appendix B)
and instantiated by GPT-4o (temperature 0.5, max 35 tokens), with the
35-token cap constraining responses to one or two sentences. An
automated GPT-4o checklist-based score (temperature 0), computed at
session close by matching the student's questions against the
scenario-specific history-taking checklist, is shown to the participant
for self-directed practice.
The platform code is publicly available at
\url{https://github.com/JINAILAB/OSCE-SimBench}.

\subsection{Real-time interaction and released dataset}

Once a participant logs in and selects a scenario, the platform hands
control to the participant, who types a free-form history-taking
question via a real-time text interface. The platform routes each
question to GPT-4o (VSP role-play), which returns a layperson-language
response constrained by behavioral rules: avoiding unsolicited
information, medical jargon, and expressions of certainty when asked
about diagnoses. All turns are appended to a session log in
chronological order, and each session is anonymized and stored as a
structured record (\texttt{session\_id}, \texttt{participant\_id} R001--R017, \texttt{clinical\_scenario}, \texttt{turn\_no}, \texttt{student\_question}, \texttt{vsp\_response}).

To date, the 17 medical-student raters who scored the static evaluation
(\S4.1) used the platform during a pilot phase, generating 49 sessions
across 27 distinct clinical scenarios between August 2025 and May 2026,
totaling approximately 1,763 student--VSP turns. The platform-collected
dialogue corpus is released as a public Hugging Face dataset (\url{https://huggingface.co/datasets/nwirandx/Korean-OSCE-history-taking-dialogues}) under CC BY-NC-SA 4.0 with a Croissant metadata file containing both Core fields
(auto-generated by Hugging Face) and Responsible-AI fields specifying
limitations, biases, intended use cases, social impact, synthetic-data
flags, and provenance. Applying the history-taking taxonomy of
\S4.3 to the platform dialogue confirms that the eleven categories
generalize beyond the static 70-prompt evaluation set: all eleven are
represented among the 1{,}614 platform utterances the scheme covers,
with the per-category distribution reported in Appendix F. The
platform thus closes the loop between static evaluation methodology and
continuously growing evaluation resource, supporting medical education
(immediate practice with feedback) and providing a privacy-compliant
dataset pipeline for downstream LLM-VSP research.

\section{Discussion}
\textbf{Limitations.} This study has several limitations. First,
the human-expert reference responses were composed by a single
OSCE-faculty author. This approach kept the reference internally
consistent across the three cases, but limits its representativeness of
OSCE educators and trained standardized patients more broadly. Where a
model did not differ significantly from the reference, that result is
therefore specific to this expert. Raters were also not asked to judge
whether a response was human-authored, so whether they could have
distinguished virtual from human responses is not recoverable from these
data. Second,
the burden of a fully-crossed human rating constrained the rater pool to
17 senior medical students and the LLM panel to four models. Several
recently released open-source frontier models that postdated the rating
window were therefore excluded from the comparison. Third, the simulator
is text-only and cannot reproduce prosody, facial expression, or
gesture, so encounter-level properties such as Engagement and
Comprehensibility may behave differently in spoken settings than the
values reported here. Fourth, the three scenarios sample only a portion
of the OSCE assessment. Korean OSCE/CPX case banks comprise more than
fifty case classes, and specialized settings such as psychiatry and
pediatrics are not represented in the present set.

\textbf{Future directions.} Several extensions follow from the present work.
First, broadening the human-reference pool by including several OSCE
educators across specialties, trained standardized patients, and more
recent commercial and open-source LLMs would strengthen the
utterance-level gold standard and improve the precision of any
comparison drawn against it. Second, the 7,480 fully-crossed multi-rater
scores released here can serve as a validation set for developing
LLM-as-a-judge evaluation pipelines, which could substantially reduce
rater burden if validated against human ratings. Third, extending the
platform beyond text to speech, video, or VR-based simulators would
yield richer student--VSP encounters and engagement signals closer to
those of a real OSCE.

In summary, this study introduces a multi-rater human-evaluation framework, a persona-injection prompt set, and a deployed VSP platform that together cover scenario authoring, evaluation, and continuous data generation for LLM-based clinical dialogue. The framework supports both medical-education practice and downstream LLM evaluation, and is released to facilitate further research on virtual standardized patients in OSCE-based training settings. Establishing the educational effect of VSP-based OSCE practice will require a controlled comparative study of learning outcomes.

\begin{ack}
This work was supported by the Bio\&Medical Technology Development Program of
the National Research Foundation (NRF), funded by the Korean government
(MSIT) (RS-2023-00220408), and by the Korea Health Technology R\&D Project
through the Korea Health Industry Development Institute (KHIDI), funded by the
Ministry of Health \& Welfare, Republic of Korea (RS-2025-02307233).
Additional support was provided by the Center for Artificial Intelligence in
Healthcare at Seoul National University Bundang Hospital (computing
infrastructure and technical advice), by the NAVER Cloud team (API credits and
access to HyperCLOVA X), and by an unrestricted donation from Dr. Sang-Gu Lee,
director of Seoul Top Internal Medicine Clinic. None of these funders, donors,
or providers had any role in the study design, the evaluation protocol, the
analysis, or the interpretation of the results. The authors declare no
competing interests.
\end{ack}


\newpage
\section*{NeurIPS Paper Checklist}

\begin{enumerate}

\item {\bf Claims}
    \item[] Question: Do the main claims made in the abstract and introduction accurately reflect the paper's contributions and scope?
    \item[] Answer: \answerYes{} 
    \item[] Justification: The abstract and \S1 list three explicit contributions (multi-rater human evaluation, question-stratified evaluation, deployed VSP platform); each is directly demonstrated in \S5 (results), \S6 (platform), and the Appendices.

\item {\bf Limitations}
    \item[] Question: Does the paper discuss the limitations of the work performed by the authors?
    \item[] Answer: \answerYes{} 
    \item[] Justification: \S7 explicitly enumerates three limitations: single-author human reference, constrained scale (n = 17 raters), and text-only modality.

\item {\bf Theory assumptions and proofs}
    \item[] Question: For each theoretical result, does the paper provide the full set of assumptions and a complete (and correct) proof?
    \item[] Answer: \answerNA{} 
    \item[] Justification: The paper presents an empirical evaluation and dataset release; no theoretical results are claimed.

    \item {\bf Experimental result reproducibility}
    \item[] Question: Does the paper fully disclose all the information needed to reproduce the main experimental results of the paper to the extent that it affects the main claims and/or conclusions of the paper (regardless of whether the code and data are provided or not)?
    \item[] Answer: \answerYes{} 
    \item[] Justification: \S3--\S4 specify all evaluation protocol details (LLM model identities, persona-injection prompt template, rater calibration procedure, statistical model). The full persona-injection prompt is reproduced in Appendix~B; the response corpus and rater scores are provided in Appendices~C and~D.

\item {\bf Open access to data and code}
    \item[] Question: Does the paper provide open access to the data and code, with sufficient instructions to faithfully reproduce the main experimental results, as described in supplemental material?
    \item[] Answer: \answerYes{} 
    \item[] Justification: All static-evaluation materials (350-response corpus, 7,480 rater scores, input-prompt set with annotations, persona-injection prompt template) are provided in full in Appendices B--E. The platform-collected dialogue corpus is released as a public Hugging Face dataset under CC BY-NC-SA 4.0 with Croissant Core and Responsible-AI metadata; the platform code is publicly available at \url{https://github.com/JINAILAB/OSCE-SimBench}.

\item {\bf Experimental setting/details}
    \item[] Question: Does the paper specify all the training and test details (e.g., data splits, hyperparameters, how they were chosen, type of optimizer) necessary to understand the results?
    \item[] Answer: \answerYes{} 
    \item[] Justification: \S3 specifies inference parameters (temperature, top-$p$, max tokens, single response per query); \S4.4 specifies the statistical model (LMM with random effects for Rater and Utterance$\mid$Case, REML estimation); \S5 reports 95\% bootstrap CIs and 95\% CIs for ICC.

\item {\bf Experiment statistical significance}
    \item[] Question: Does the paper report error bars suitably and correctly defined or other appropriate information about the statistical significance of the experiments?
    \item[] Answer: \answerYes{} 
    \item[] Justification: All effect estimates are reported with 95\% confidence intervals; bootstrap CIs are used for descriptive means; F-distribution method for ICC; LMM contrasts use Wald-style 95\% CIs.

\item {\bf Experiments compute resources}
    \item[] Question: For each experiment, does the paper provide sufficient information on the computer resources (type of compute workers, memory, time of execution) needed to reproduce the experiments?
    \item[] Answer: \answerNA{} 
    \item[] Justification: The paper evaluates externally hosted LLM endpoints; no model training or fine-tuning was performed. Inference incurred negligible compute cost (single response per prompt across 350 prompts $\times$ 4 LLMs).
    
\item {\bf Code of ethics}
    \item[] Question: Does the research conducted in the paper conform, in every respect, with the NeurIPS Code of Ethics \url{https://neurips.cc/public/EthicsGuidelines}?
    \item[] Answer: \answerYes{} 
    \item[] Justification: The research conforms to the NeurIPS Code of Ethics: human raters were recruited voluntarily under written informed consent and IRB approval; compensation was at the institution's standard research-participation rate.

\item {\bf Broader impacts}
    \item[] Question: Does the paper discuss both potential positive societal impacts and negative societal impacts of the work performed?
    \item[] Answer: \answerYes{} 
    \item[] Justification: \S1 and \S6 describe positive societal impact (cost-effective OSCE training resource accessible across institutions). Negative-impact considerations (e.g., over-reliance on automated VSPs, language-coverage limitations) are addressed in \S7 limitations and the Datasheet (Appendix~A).
    
\item {\bf Safeguards}
    \item[] Question: Does the paper describe safeguards that have been put in place for responsible release of data or models that have a high risk for misuse (e.g., pre-trained language models, image generators, or scraped datasets)?
    \item[] Answer: \answerNA{} 
    \item[] Justification: The released artefacts are evaluation data (rater scores, response transcripts) and dialogue logs; no high-risk pretrained models or generative checkpoints are released.

\item {\bf Licenses for existing assets}
    \item[] Question: Are the creators or original owners of assets (e.g., code, data, models), used in the paper, properly credited and are the license and terms of use explicitly mentioned and properly respected?
    \item[] Answer: \answerYes{} 
    \item[] Justification: Every cited LLM, dataset, and external software package is named in the manuscript with its source and licensing terms summarized in the Datasheet (Appendix~A); the released platform-collected dataset is licensed under CC BY-NC-SA 4.0.

\item {\bf New assets}
    \item[] Question: Are new assets introduced in the paper well documented and is the documentation provided alongside the assets?
    \item[] Answer: \answerYes{} 
    \item[] Justification: The platform-collected dialogue corpus is documented via a Datasheet for Datasets (Appendix~A) and a Croissant metadata file with Core and Responsible-AI fields; intended use, limitations, and biases are described.

\item {\bf Crowdsourcing and research with human subjects}
    \item[] Question: For crowdsourcing experiments and research with human subjects, does the paper include the full text of instructions given to participants and screenshots, if applicable, as well as details about compensation (if any)? 
    \item[] Answer: \answerYes{} 
    \item[] Justification: Seventeen senior medical-student raters were recruited via an open invitation circulated through the medical-education office; they completed a 60-minute calibration session before scoring; participants received compensation at the institution's standard student research-participation rate. Rating instructions (the Likert agreement statements presented to raters) are reproduced in Table~1 and Appendix~C.

\item {\bf Institutional review board (IRB) approvals or equivalent for research with human subjects}
    \item[] Question: Does the paper describe potential risks incurred by study participants, whether such risks were disclosed to the subjects, and whether Institutional Review Board (IRB) approvals (or an equivalent approval/review based on the requirements of your country or institution) were obtained?
    \item[] Answer: \answerYes{} 
    \item[] Justification: All scenarios, the input-prompt set, and the rating protocol were reviewed and approved by the Institutional Review Board of Dankook University Hospital prior to data collection. Participation was voluntary, written informed consent was obtained, and the platform-collected dataset is released under research-participation and personal-information consent.

\item {\bf Declaration of LLM usage}
    \item[] Question: Does the paper describe the usage of LLMs if it is an important, original, or non-standard component of the core methods in this research? Note that if the LLM is used only for writing, editing, or formatting purposes and does \emph{not} impact the core methodology, scientific rigor, or originality of the research, declaration is not required.
    \item[] Answer: \answerYes{} 
    \item[] Justification: LLMs (GPT-4o, Claude 3.5 Sonnet, Llama-3.1 70B, HyperCLOVA X) constitute the experimental subjects of this evaluation; their roles and configurations are fully described in \S3.3 (response generation) and \S6.1 (platform deployment).

\end{enumerate}

\appendix

\raggedbottom

\setlength{\intextsep}{\baselineskip}
\setlength{\textfloatsep}{\baselineskip}
\setlength{\floatsep}{\baselineskip}
\captionsetup{aboveskip=0.5\baselineskip, belowskip=0.5\baselineskip}

\setcounter{figure}{0}
\setcounter{table}{0}
\renewcommand{\thefigure}{\thesection\arabic{figure}}
\renewcommand{\thetable}{\thesection\arabic{table}}
\makeatletter
\@addtoreset{figure}{section}
\@addtoreset{table}{section}
\makeatother

\section{Datasheet for Datasets}

This Datasheet documents the platform-collected dialogue corpus released
on Hugging Face. It follows the seven-section template of Gebru et al.
(2018). The static evaluation materials referenced in the main paper
(the 350-response corpus, 7,480 rater scores, and persona-injection
prompt template) are provided in Appendices B--F and are not part of the
publicly hosted dataset described here.

\subsection*{A.1 Motivation}

\textbf{For what purpose was the dataset created?} To support research on
LLM-based VSPs in Korean OSCE-style medical education by capturing
real-time student--VSP dialogue collected on the deployed VSP-OSCE
training platform.

\textbf{Who created the dataset?} The dataset was created by the authors of
this work. The Likert ratings were contributed by 17 senior medical-student
raters recruited with informed consent; the raters are not authors of this
work.

\textbf{Who funded the creation of the dataset?} The Bio\&Medical Technology
Development Program of the National Research Foundation (NRF) funded by the
Korean government (MSIT) (RS-2023-00220408), and the Korea Health Technology
R\&D Project through the Korea Health Industry Development Institute (KHIDI),
funded by the Ministry of Health \& Welfare, Republic of Korea
(RS-2025-02307233).

\textbf{Any other comments?} Created under IRB-approved informed consent at
Dankook University Hospital.

\subsection*{A.2 Composition}

\textbf{What do the instances that comprise the dataset represent?} Single
turn-pairs from real-time student--VSP encounters: one student question
(human-typed in Korean) and one VSP response (GPT-4o under persona
injection).

\textbf{How many instances are there in total?} 49 sessions across 27
scenarios, \\~{}1,763 turn-pairs from 17 medical-student participants.

\textbf{Does the dataset contain all possible instances or is it a sample?}
All turn-pairs generated by the cohort during 2025-08 to 2026-05; not a
subsample.

\textbf{What data does each instance consist of?} Fields: session\_id,
participant\_id (R001--R017), clinical\_scenario, turn\_no,
student\_question, vsp\_response. Per-participant summary statistics
provided separately.

\textbf{Is there a label or target associated with each instance?} None. Raw
dialogue.

\textbf{Is any information missing from individual instances?} No structural
fields missing; a few sessions terminated early but all collected turns
are retained.

\textbf{Are relationships between individual instances made explicit?} Yes
--- via session\_id (groups turns) and participant\_id (links sessions).

\textbf{Are there recommended data splits?} None.

\textbf{Are there any errors, sources of noise, or redundancies?} Occasional
typing errors and rare role-inconsistent VSP responses; no redundancies.

\textbf{Is the dataset self-contained, or does it link to external
resources?} Self-contained; persona-injection prompt provided as
accompanying file (also in Appendix B).

\textbf{Does the dataset contain confidential data?} No. Identifiers replaced
with R001--R017; mapping destroyed.

\textbf{Does the dataset contain offensive, insulting, or harmful content?}
No.

\textbf{Does the dataset relate to people?} Yes --- 17 medical-student
participants at Korean medical schools (clinical-clerkship level), all
anonymized.

\textbf{Does the dataset identify any subpopulations?} Year-of-training and
aggregate metrics released; no fine-grained demographics or PHI.

\textbf{Is it possible to identify individuals from the dataset?} No.

\textbf{Does the dataset contain sensitive data?} No clinical PHI; scenarios
are synthetic OSCE cases.

\subsection*{A.3 Collection process}

\textbf{How was the data acquired?} Via the VSP-OSCE platform: GPT-4.1
(T=0.2) for scenario generation, GPT-4o (T=0) for automated evaluation,
GPT-4o (T=0.5, max 35 tokens) for patient role-play under persona
injection --- all with seed=42. Dataset records the role-play output.

\textbf{What mechanisms or procedures were used to collect the data?}
Real-time text-based interaction logged automatically; identifiers
anonymized post-session.

\textbf{If the dataset is a sample from a larger population, what was the
sampling strategy?} Convenience sample of 17 medical-student
volunteers.

\textbf{Who collected the data, and how were they compensated?} Research team
via the platform; participants paid at standard student
research-participation rate.

\textbf{Over what timeframe was the data collected?} 2025-08-01 to
2026-05-03.

\textbf{Were any ethical review processes conducted?} Yes --- IRB approval at
Dankook University Hospital.

\textbf{Was the data collected directly from individuals?} Yes.

\textbf{Were individuals notified about the collection?} Yes --- written
information including purpose, anonymization, and right to withdraw.

\textbf{Did the individuals consent to the collection and use of their data?}
Yes --- research-participation and personal-information consent in
writing.

\textbf{Was a mechanism provided to revoke consent?} Yes --- withdrawn
participants are removed from subsequent dataset releases.

\textbf{Has an analysis of the potential impact of the dataset and its use on
data subjects been conducted?} IRB review served as the ethics
analysis; anonymization is the principal mitigation.

\textbf{Does the dataset contain synthetic data?} Yes --- VSP-side answers
(\\~{}1,763 turns) are GPT-4o-generated under persona injection. Student
questions are human-typed.

\textbf{Was the dataset derived from another dataset?} No external dataset.
27 scenarios authored by the research team from an 80-symptom $\times$
3-profile internal database; verified by a diagnostic-judgment model +
expert review.

\subsection*{A.4 Preprocessing, cleaning, and labeling}

\textbf{Was any preprocessing or cleaning of the data performed?}
Anonymization: identifiers replaced with R001--R017; in-text real names
substituted; original mapping destroyed.

\textbf{Was the raw data saved in addition to the preprocessed data?} No.

\textbf{Is the software used to preprocess the data available?} Yes ---
released with the dataset.

\textbf{Were any annotations or labels added beyond the raw dialogue?} No
additional labels beyond the structured-record fields.

\subsection*{A.5 Uses}

\textbf{Has the dataset been used for any tasks already?} Yes --- referenced
in \S6 of the present paper.

\textbf{Is there a repository linking to all uses of the dataset?} No central
repository of downstream uses is maintained. Known uses will be listed on the
dataset card at \url{https://huggingface.co/datasets/nwirandx/Korean-OSCE-history-taking-dialogues}.

\textbf{What other tasks could the dataset be used for?} LLM-VSP development;
OSCE training research; clinical-dialogue benchmarking; post-training
and preference optimization of domain-specialized LLMs.

\textbf{Is there anything about the composition or collection that might
affect future uses?} Korean-language; senior medical students (not
clinicians); text-only; 27 internal-medicine scenarios. Validity
established for LLM-VSP research, OSCE training, and multi-rater
methodology; NOT for clinical decision-making, diagnosis, non-Korean
contexts, or standalone training.

\textbf{Are there tasks for which the dataset should not be used?} Clinical
decision-making; patient diagnosis; standalone medical training without
supervision.

\textbf{Other comments on social impact?} Positive: the dataset enables
cost-effective, reproducible OSCE practice for medical students and
supports research on LLM-based clinical dialogue agents. Negative:
overreliance on LLM patient simulation could displace human
standardized-patient programs, and high LLM scores could be misread as
clinical competence rather than role-play fidelity. Mitigations:
explicit unsuitable-use-case statements, IRB-approved consent, and full
participant anonymization.

\subsection*{A.6 Distribution}

\textbf{Will the dataset be distributed to third parties outside the research
team?} Yes --- public Hugging Face dataset under IRB-approved consent.

\textbf{How will the dataset be distributed?} Hugging Face at \url{https://huggingface.co/datasets/nwirandx/Korean-OSCE-history-taking-dialogues} with a Croissant Core + RAI metadata file.

\textbf{When will the dataset be distributed?} On paper acceptance; reviewers
may access via the URL above.

\textbf{Will the dataset be released under a copyright or other IP license?}
CC BY-NC-SA 4.0 (attribution + non-commercial + share-alike).

\textbf{Are there other restrictions on the use of the dataset?} No
re-identification; no use for clinical decision-making, diagnosis, or
standalone medical training.

\textbf{Are there export controls or other regulatory restrictions?} None
known.

\subsection*{A.7 Maintenance}

\textbf{Who is supporting / hosting / maintaining the dataset?} The authors of
this work; hosted on Hugging Face.

\textbf{How can the owner / curator / manager be contacted?}
\texttt{yongjin.neph@gmail.com}.

\textbf{Is there an erratum?} None at submission; tracked via GitHub Issues +
dataset versioning.

\textbf{Will the dataset be updated?} Yes --- new sessions appended as
cohorts use the platform.

\textbf{Are there limits on retention of data associated with individuals?}
No technical limit; withdrawn participants are removed.

\textbf{Will older versions of the dataset be supported?} Yes ---
version-tagged on Hugging Face.

\textbf{Is there a mechanism for others to extend or build on the dataset?}
Yes --- PRs on GitHub or extension issues on Hugging Face; MIT-licensed
code allows third-party deployment subject to local IRB approval.

\textbf{Other comments?} Part of a cyclic pipeline (construction $\to$ evaluation
$\to$ training $\to$ data generation) for sustainable virtual OSCE education and
LLM post-training.
\section{Persona-Injection Prompt}

The persona-injection prompt used for all three OSCE cases follows a single shared template (English translation shown below; the Korean originals are released alongside the dataset). The three case-specific instantiations differ only in the chief-complaint phrase, patient identifier, and scenario body; the surrounding instructions are otherwise identical, with one optional case-specific behavioral directive used only for the pregnancy case. We illustrate the template with one fully-instantiated worked example (hematuria); the remaining two case instantiations follow the same structure.

\subsection*{B.1 Unified template (English translation)}

\begin{quote}\small
The following information is the prior medical record of a patient presenting with [CHIEF\_COMPLAINT] at the clinical-skills licensure examination. Based on this information, conduct a conversation between the [PATIENT\_DESCRIPTOR] and a student-doctor.

\medskip
[SCENARIO\_BODY --- case-specific clinical narrative covering patient demographics, chief-complaint trajectory, associated symptoms, past medical / family / social history, and patient concerns; an explicit list of negative findings is included to keep VSP responses scenario-grounded.]

\medskip
You are now the simulated patient [PATIENT\_NAME]. Please answer the questions I ask in fewer than 30 words. All answers must be complete sentences. Do not answer content that has not been asked. [OPTIONAL\_CASE\_SPECIFIC\_DIRECTIVE]
\end{quote}

\subsection*{B.2 Per-case variable assignments}

\begin{table}[H]
  \centering\footnotesize
  \setlength{\tabcolsep}{4pt}
  \caption{Per-case variable assignments for the unified persona-injection template (English).}
  \label{tab:b1}
  \begin{tabular}{p{0.19\linewidth}p{0.19\linewidth}p{0.32\linewidth}p{0.19\linewidth}}
    \toprule
    \textbf{Variable} & \textbf{Hematuria} & \textbf{Fever (pregnancy)} & \textbf{Chronic cough} \\
    \midrule
    CHIEF\_\hspace{0pt}COMPLAINT & red urine & fever (pregnant patient) & chronic cough \\
    PATIENT\_\hspace{0pt}DESCRIPTOR & patient with red urine & pregnant patient & patient with chronic cough \\
    PATIENT\_\hspace{0pt}NAME & Mr.\ Jung Jin-su (28-year-old male) & Ms.\ Lee Su-jin (28-year-old female, 20-week gestation) & Ms.\ Lee Yun-jung (55-year-old female) \\
    OPTIONAL\_\hspace{0pt}DIRECTIVE & --- & When the student-doctor explains that antibiotic treatment is needed, the patient should explicitly ask whether antibiotics may be harmful to the unborn child. & --- \\
    \bottomrule
  \end{tabular}
\end{table}

\subsection*{B.3 Worked example --- hematuria case (English translation)}

\begin{quote}\small
The following information is the prior medical record of a patient presenting with red urine at the clinical-skills licensure examination. Based on this information, conduct a conversation between the patient with red urine and a student-doctor.

\medskip
Mr.\ Jung Jin-su is a 28-year-old male. He came to the hospital because of red urine that started three days ago. He had cold symptoms beginning a week ago, and on the day before the red urine appeared his temperature rose to 39\,\textdegree{}C with associated muscle aches. The muscle aches and fever were most severe on the evening four days ago; the next morning (three days ago) he began passing red urine. The color was dark reddish-brown --- initially deep reddish-brown, gradually lightening to dark amber today. He has not previously been told he had hematuria, proteinuria, or kidney disease. He takes no regular medications, herbal supplements, or health products. He has been told his blood pressure is somewhat elevated but did not consider it serious and has never visited a hospital; he does not remember the exact value. This is the first time his urine has appeared abnormal. He had no abdominal pain, flank pain, or pain at the urethral tip while passing red urine. He thinks there might have been a slight foam in the urine, but he is not sure.

\medskip
Otherwise, he has not felt weak, his weight has not changed, and he denies headache, dizziness, fever, chills, muscle pain, night sweats, cough, sputum, runny nose, chest pain, dyspnoea, palpitations, exertional dyspnoea, abdominal pain, nausea, vomiting, diarrhoea, constipation, haematemesis, melaena, or haematochezia. His appetite is normal.

\medskip
Mr.\ Jung has considered himself healthy and has rarely visited hospitals. At his early-twenties military medical examination and last year's workplace check-up, no abnormality other than slightly elevated blood pressure was noted. He has smoked about one pack a day since his military service. He drinks beer one or two glasses only at company gatherings. He has worked as an office worker for two years and undergoes annual workplace health check-ups. He used to exercise at the gym for one hour each day after work but has not gone since catching the cold a week ago. He prefers meat, eats a slightly salty diet, and does not drink much water. He works 9 AM to 6 PM and lives nearby with his family. He has a girlfriend with whom he has sex about once a week, always using a condom. He graduated from a four-year university and served in the military during his sophomore year. His father has hypertension and is on antihypertensive medication; his mother and one older brother have no known health problems.

\medskip
Mr.\ Jung is worried about the red urine. After searching online, he found that it is called hematuria and may indicate stones or cancer; he is very concerned that he may have a serious illness. He wonders whether tests will be done today and whether he will see the results, and is also curious whether hospitalisation or detailed examinations will be required.

\medskip
You are now the simulated patient Mr.\ Jung Jin-su. Please answer the questions I ask in fewer than 30 words. All answers must be complete sentences. Do not answer content that has not been asked.
\end{quote}

\section{Translated Generated Dialogues}

This appendix provides English translations of the full 350-response
evaluation corpus referenced in the main paper: 70 input prompts $\times$ 5
sources (four LLMs and the human-expert reference) across the three OSCE
cases (hematuria, fever in pregnancy, chronic cough). For each input
prompt, the translated source-doctor question is followed by the five
sources' translated responses. Korean originals are released alongside
the static-evaluation materials. Speaker labels are abbreviated
throughout this appendix: \emph{Human} denotes the human-expert
reference, and \emph{Claude}, \emph{Llama}, and \emph{HyperCLOVA} denote
Claude 3.5 Sonnet, Llama-3.1 70B, and HyperCLOVA X, respectively.

\subsection*{Case: Hematuria (red urine) \textnormal{\textit{(translated from Korean)}}}

\textbf{Q1.} Hello, I'm medical student Kim Si-jun. May I have your name and
age, please?

\par\noindent$\bullet$~Human: Hello, my name is Jung Jin-su, and I'm 28 years old.

\par\noindent$\bullet$~Claude: Hello, my name is Jung Jin-su, 28 years old.

\par\noindent$\bullet$~GPT-4o: Hello, I'm Jung Jin-su, 28 years old.

\par\noindent$\bullet$~HyperCLOVA: Jung Jin-su: Hello, my name is Jung Jin-su, 28 years old.

\par\noindent$\bullet$~Llama: Yes, my name is Jung Jin-su, 28 years old.

\textbf{Q2.} Hello, Mr. Jung. What brings you in today?

\par\noindent$\bullet$~Human: My urine suddenly turned red, so I came in.

\par\noindent$\bullet$~Claude: I came to the hospital because of red urine that started three
days ago.

\par\noindent$\bullet$~GPT-4o: Hello, I'm here because red urine has been appearing for the
past three days.

\par\noindent$\bullet$~HyperCLOVA: Yes. My urine has been red for three days, so I came in.

\par\noindent$\bullet$~Llama: I've had red urine since three days ago.

\textbf{Q3.} What was the color of the red urine like --- closer to brown,
or pinkish?

\par\noindent$\bullet$~Human: It seemed close to a dark brown.

\par\noindent$\bullet$~Claude: Initially it was deep reddish-brown, gradually lightening to
about dark amber today.

\par\noindent$\bullet$~GPT-4o: At first it was deep reddish-brown, gradually lightening to
about dark amber today.

\par\noindent$\bullet$~HyperCLOVA: On the first day it was very deep reddish-brown. Now it's
lighter --- about dark amber.

\par\noindent$\bullet$~Llama: Initially it was deep reddish-brown, gradually lightening to
about dark amber today.

\textbf{Q4.} Was there any blood clot or stone in the urine?

\par\noindent$\bullet$~Human: No, there wasn't anything like that.

\par\noindent$\bullet$~Claude: I didn't see any clots or stones. There seemed to be a slight
foam in the urine.

\par\noindent$\bullet$~GPT-4o: No, there were no clots or stones.

\par\noindent$\bullet$~HyperCLOVA: No, nothing like that. There seemed to be slight foam in
the urine.

\par\noindent$\bullet$~Llama: No, there were no clots or stones.

\textbf{Q5.} Have you ever had red urine before? Was hematuria ever mentioned
in school or workplace check-ups?

\par\noindent$\bullet$~Human: No, this is the first time. I've never been told about it on
check-ups either.

\par\noindent$\bullet$~Claude: This is the first time I've had red urine. I've never been
told about hematuria on check-ups.

\par\noindent$\bullet$~GPT-4o: No, I've never been told about hematuria, proteinuria, or
kidney disease, and this is the first time the urine has appeared
abnormal.

\par\noindent$\bullet$~HyperCLOVA: No, this kind of urine is the first time. I haven't heard
about hematuria on check-ups either.

\par\noindent$\bullet$~Llama: No, I've never had red urine before, and the school and
workplace check-ups never mentioned hematuria either.

\textbf{Q6.} That must have been distressing. Around the time the red urine
appeared, did you have a cold or do strenuous exercise?

\par\noindent$\bullet$~Human: I had a cold for about a week. The red urine appeared while I
was at the peak of being sick.

\par\noindent$\bullet$~Claude: Yes, I had cold symptoms before the red urine appeared, and my
fever went up to 39 \textdegree{}C. I didn't exercise.

\par\noindent$\bullet$~GPT-4o: Yes, the day before the red urine I had a fever up to 39 \textdegree{}C
with muscle aches and cold symptoms. I haven't exercised since the cold
started a week ago.

\par\noindent$\bullet$~HyperCLOVA: Yes, I happened to have a cold, with severe fever and
muscle aches in particular, so I took antipyretics.

\par\noindent$\bullet$~Llama: Yes, I had a cold. Cold symptoms started seven days ago, and
the day before the red urine my fever rose to 39 \textdegree{}C with muscle aches.

\textbf{Q7.} When you had the red urine, did you have any flank pain or
abdominal pain?

\par\noindent$\bullet$~Human: There was no abdominal or flank pain.

\par\noindent$\bullet$~Claude: No abdominal pain, flank pain, or pain at the urethral tip
when I had red urine.

\par\noindent$\bullet$~GPT-4o: No, there was no abdominal pain, flank pain, or pain at the
urethral tip with the red urine.

\par\noindent$\bullet$~HyperCLOVA: No, there was no pain at all when passing urine.

\par\noindent$\bullet$~Llama: No, there was no flank pain or abdominal pain.

\textbf{Q8.} Have you had any swelling of the legs or face?

\par\noindent$\bullet$~Human: No swelling.

\par\noindent$\bullet$~Claude: No swelling of the legs or face. My weight has not changed.

\par\noindent$\bullet$~GPT-4o: No, my weight hasn't changed and I haven't felt weak or had
any swelling.

\par\noindent$\bullet$~HyperCLOVA: No weight change, and no particular swelling.

\par\noindent$\bullet$~Llama: No, I haven't had any swelling of the legs or face.

\textbf{Q9.} Have you had joint pain anywhere or red spots on the skin?

\par\noindent$\bullet$~Human: No, nothing like that.

\par\noindent$\bullet$~Claude: No joint pain or red spots on the skin.

\par\noindent$\bullet$~GPT-4o: No, there have been no muscle pain or red skin spots.

\par\noindent$\bullet$~HyperCLOVA: No, only the urine color was strange.

\par\noindent$\bullet$~Llama: No, I've had no joint pain or red skin spots.

\textbf{Q10.} When you passed red urine, did you have pain at the end of
urination, or any discomfort?

\par\noindent$\bullet$~Human: No, there was no discomfort when urinating.

\par\noindent$\bullet$~Claude: I had no pain or discomfort at all when urinating.

\par\noindent$\bullet$~GPT-4o: No, no pain at the urethral tip or any other discomfort when
urinating.

\par\noindent$\bullet$~HyperCLOVA: No, that wasn't the case. I was actually surprised at how
normal it felt.

\par\noindent$\bullet$~Llama: No, no pain at the urethral tip or discomfort with urination.

\textbf{Q11.} Have you been urinating very frequently lately, or feeling that
you can't fully empty your bladder?

\par\noindent$\bullet$~Human: No, I haven't had those symptoms.

\par\noindent$\bullet$~Claude: No increase in frequency or sense of incomplete emptying. No
urinary discomfort.

\par\noindent$\bullet$~GPT-4o: No, no increased frequency or sense of incomplete emptying.

\par\noindent$\bullet$~HyperCLOVA: I don't drink much water, so I don't urinate often, and
I haven't felt incomplete emptying.

\par\noindent$\bullet$~Llama: No, I haven't had increased frequency or incomplete emptying
recently.

\textbf{Q12.} Have you noticed any foam in your urine?

\par\noindent$\bullet$~Human: Yes, I did notice some slight foam.

\par\noindent$\bullet$~Claude: I'm not sure, but there seemed to be a slight foam in the
urine.

\par\noindent$\bullet$~GPT-4o: I'm not sure, but there seemed to be a slight foam in the
urine.

\par\noindent$\bullet$~HyperCLOVA: Not sure, but I think there was a little.

\par\noindent$\bullet$~Llama: Yes, there seemed to be a slight foam in the urine.

\textbf{Q13.} Are you on any regular medication, or did you take anything for
the cold recently?

\par\noindent$\bullet$~Human: When the cold was severe I bought Tylenol from the pharmacy and
took it for a few days. I've stopped now.

\par\noindent$\bullet$~Claude: No regular medications. I bought Tylenol from the pharmacy for
the cold.

\par\noindent$\bullet$~GPT-4o: No regular medication. I took Tylenol because of the cold
symptoms.

\par\noindent$\bullet$~HyperCLOVA: I take no regular medications. I just bought and took
Tylenol when I had the cold.

\par\noindent$\bullet$~Llama: Yes, I took Tylenol that I bought from the pharmacy for the
cold.

\textbf{Q14.} Do you have any pre-existing kidney disease or have you had any
major surgery?

\par\noindent$\bullet$~Human: No, nothing like that.

\par\noindent$\bullet$~Claude: No kidney disease or major surgery. I've considered myself
healthy.

\par\noindent$\bullet$~GPT-4o: No, I've never been told about hematuria, proteinuria, or
kidney disease, and I've never had major surgery.

\par\noindent$\bullet$~HyperCLOVA: No, I'm very healthy.

\par\noindent$\bullet$~Llama: No, no kidney disease or major surgery.

\textbf{Q15.} Does any family member have hematuria or kidney disease, or any
urological cancer?

\par\noindent$\bullet$~Human: No, no family members with such illnesses, as far as I know.

\par\noindent$\bullet$~Claude: No family member has been diagnosed with hematuria, kidney
disease, or urological cancer. My father has high blood pressure.

\par\noindent$\bullet$~GPT-4o: No, no family members have been diagnosed with hematuria,
kidney disease, or urological cancer.

\par\noindent$\bullet$~HyperCLOVA: I know my father has high blood pressure and takes
medication. Other than that, none.

\par\noindent$\bullet$~Llama: No, no family members have hematuria, kidney disease, or
urological cancer.

\textbf{Q16.} Do you smoke?

\par\noindent$\bullet$~Human: Yes, about a pack a day.

\par\noindent$\bullet$~Claude: Yes, since my military service I've smoked about one pack a
day.

\par\noindent$\bullet$~GPT-4o: Yes, I usually smoke about one pack a day.

\par\noindent$\bullet$~HyperCLOVA: Yes. I started in the military and smoke about one pack a
day.

\par\noindent$\bullet$~Llama: Yes, since my military service I've smoked about one pack a
day.

\textbf{Q17.} Do you drink alcohol? If so, how much?

\par\noindent$\bullet$~Human: I drink only at company gatherings, just a little.

\par\noindent$\bullet$~Claude: I only drink one or two glasses of beer at company gatherings.

\par\noindent$\bullet$~GPT-4o: Yes, I drink only at company gatherings, one or two glasses of
beer.

\par\noindent$\bullet$~HyperCLOVA: Hardly. Just one or two glasses of beer at company
gatherings.

\par\noindent$\bullet$~Llama: Yes, I drink one or two glasses of beer only at company
gatherings.

\textbf{Q18.} I'd like to ask carefully --- have you had any sexual contact
with a stranger recently, or felt any pain after intercourse?

\par\noindent$\bullet$~Human: No, nothing like that.

\par\noindent$\bullet$~Claude: I only have sex with my girlfriend, about once a week. No
contact with strangers and no pain.

\par\noindent$\bullet$~GPT-4o: No, no recent sexual contact with strangers, and no pain after
intercourse.

\par\noindent$\bullet$~HyperCLOVA: I only have intercourse with my girlfriend, always with a
condom. And nothing like that has happened.

\par\noindent$\bullet$~Llama: No, I haven't had any recent sexual contact with strangers,
and no pain.

\textbf{Q19.} Is your urine stream weak, or do you have to strain to urinate?

\par\noindent$\bullet$~Human: No, neither.

\par\noindent$\bullet$~Claude: No, no weak stream and no straining required.

\par\noindent$\bullet$~GPT-4o: No, my urine stream isn't weak and I don't have to strain.

\par\noindent$\bullet$~HyperCLOVA: No, it's the same as usual.

\par\noindent$\bullet$~Llama: No, my stream isn't weak and I don't have to strain.

\textbf{Q20.} Thank you. To summarize: you had a cold a week ago, and at the
peak of the fever you noticed reddish-brown urine, with no pain on
urination, abdominal pain, or flank pain. Is that correct?

\par\noindent$\bullet$~Human: Yes, that's correct.

\par\noindent$\bullet$~Claude: Yes, that's right. After cold symptoms and high fever I had
reddish-brown urine, with no pain.

\par\noindent$\bullet$~GPT-4o: Yes, that's right. A week ago I had cold symptoms; at the
peak of fever I had reddish-brown urine and no pain or flank pain on
urination.

\par\noindent$\bullet$~HyperCLOVA: Yes, exactly. That's what I said.

\par\noindent$\bullet$~Llama: Yes, that's correct.

\textbf{Q21.} Given the picture, this is likely post-cold-related hematuria.
Hematuria means blood from the kidneys or urinary tract has mixed into
the urine. Without pain or clots, stones are less likely; cold-related
stress on the kidney glomerulus is a more probable cause. Does that make
sense?

\par\noindent$\bullet$~Human: Yes, I understand. So what tests do I need?

\par\noindent$\bullet$~Claude: Yes, I understand. So you're saying cold-related stress
likely affected the kidneys and caused hematuria.

\par\noindent$\bullet$~GPT-4o: Yes, I understand. The explanation about cold-related stress
causing glomerular hematuria is clear. Thank you.

\par\noindent$\bullet$~HyperCLOVA: Yes, you mean it could be due to the cold, right?

\par\noindent$\bullet$~Llama: Yes, I understand.

\textbf{Q22.} For diagnosis we'll need a urinalysis, microscopy, proteinuria
check, and a blood test for kidney function. If anything is abnormal we
may need imaging or a kidney biopsy.

\par\noindent$\bullet$~Human: Alright.

\par\noindent$\bullet$~Claude: Yes, I understand. So urinalysis, blood tests, and
kidney-function tests first, with additional tests if needed.

\par\noindent$\bullet$~GPT-4o: Yes, I understand. Urinalysis, microscopy, proteinuria
testing, and blood-based kidney-function tests are needed; additional
detailed tests if abnormal.

\par\noindent$\bullet$~HyperCLOVA: Yes, I see. Can I get the tests done right away?

\par\noindent$\bullet$~Llama: Yes, I understand.

\textbf{Q23.} Lastly, since hematuria may be associated with smoking, I would
recommend quitting.

\par\noindent$\bullet$~Human: Alright. I'll quit smoking. Thank you.

\par\noindent$\bullet$~Claude: Yes, I'll consider quitting since smoking may be linked to
hematuria.

\par\noindent$\bullet$~GPT-4o: Yes, I understand the link between smoking and hematuria, and
I'll consider quitting. Thank you for the advice.

\par\noindent$\bullet$~HyperCLOVA: Yes, understood. I'll quit immediately.

\par\noindent$\bullet$~Llama: Yes, I understand. I'll quit smoking.

\subsection*{Case: Fever (in pregnancy) \textnormal{\textit{(translated from Korean)}}}

\textbf{Q1.} Hello, I'm medical student Kang Min-woo. May I have your name
and age, please?

\par\noindent$\bullet$~Human: Hello, I'm Lee Su-jin, 28 years old.

\par\noindent$\bullet$~Claude: Hello, my name is Lee Su-jin, 28 years old.

\par\noindent$\bullet$~GPT-4o: Hello, I'm Lee Su-jin, 28 years old.

\par\noindent$\bullet$~HyperCLOVA: Yes. Hello, I'm Lee Su-jin, 28 years old.

\par\noindent$\bullet$~Llama: Hello, I'm Lee Su-jin, 28 years old.

\textbf{Q2.} Hello, Ms. Lee. What brings you in today?

\par\noindent$\bullet$~Human: I came in because I had a high fever from yesterday evening.

\par\noindent$\bullet$~Claude: Since yesterday evening I've had a high fever and muscle pain
all over. My right flank especially hurts a lot.

\par\noindent$\bullet$~GPT-4o: I came in because of fever since yesterday and pain in my
right flank.

\par\noindent$\bullet$~HyperCLOVA: I came in because of fever and muscle pain all over since
yesterday evening.

\par\noindent$\bullet$~Llama: I have severe body aches from the high fever, with sharp pain
in the right flank in particular.

\textbf{Q3.} That sounds difficult. How high did the fever go, and is it
constant or intermittent?

\par\noindent$\bullet$~Human: Yesterday evening it measured up to 39 \textdegree{}C, and even after
Tylenol the fever barely came down.

\par\noindent$\bullet$~Claude: It went up to 39 \textdegree{}C on the thermometer. The fever was
continuous, and Tylenol didn't bring it down well.

\par\noindent$\bullet$~GPT-4o: The fever went up to 39 \textdegree{}C, and it seems constant.

\par\noindent$\bullet$~HyperCLOVA: It went up to 39 \textdegree{}C and was sustained continuously.

\par\noindent$\bullet$~Llama: Measured at 39 \textdegree{}C. The fever started yesterday evening and has
continued to now.

\textbf{Q4.} Did you have chills or shivering?

\par\noindent$\bullet$~Human: Yes, when it's bad I feel very cold and shivery.

\par\noindent$\bullet$~Claude: Yes, there were some chills. I felt shivery.

\par\noindent$\bullet$~GPT-4o: There were some chills.

\par\noindent$\bullet$~HyperCLOVA: Slight chills accompanied it.

\par\noindent$\bullet$~Llama: No, no cold sweats.

\textbf{Q5.} Any sore throat, cough, or sputum?

\par\noindent$\bullet$~Human: No sore throat, cough, or sputum.

\par\noindent$\bullet$~Claude: No, no sore throat, cough, or sputum.

\par\noindent$\bullet$~GPT-4o: No, no cough or sputum.

\par\noindent$\bullet$~HyperCLOVA: No cough, no sputum.

\par\noindent$\bullet$~Llama: No, no cough. No sputum either.

\textbf{Q6.} Any diarrhoea or vomiting?

\par\noindent$\bullet$~Human: No.

\par\noindent$\bullet$~Claude: No vomiting and no diarrhoea, though I did feel a bit
nauseous.

\par\noindent$\bullet$~GPT-4o: No, no diarrhoea or vomiting.

\par\noindent$\bullet$~HyperCLOVA: No diarrhoea, no vomiting.

\par\noindent$\bullet$~Llama: No, no vomiting. No diarrhoea either.

\textbf{Q7.} Any pain in the flank or lower abdomen?

\par\noindent$\bullet$~Human: Yes, the right flank in particular hurts.

\par\noindent$\bullet$~Claude: Yes, sharp pain in the right flank when moving. The lower
abdomen isn't particularly painful.

\par\noindent$\bullet$~GPT-4o: Yes, sharp pain in the right flank.

\par\noindent$\bullet$~HyperCLOVA: Sharp pain in the right flank when moving.

\par\noindent$\bullet$~Llama: No, the flank pain has eased somewhat, and no lower-abdominal
pain.

\textbf{Q8.} What is the right flank pain like? Please describe in detail.

\par\noindent$\bullet$~Human: It's fine when still, but when I try to move, the right flank
stabs sharply.

\par\noindent$\bullet$~Claude: The right flank pain is worse on movement, sharp and stabbing.
After Tylenol it improved a bit but is still uncomfortable.

\par\noindent$\bullet$~GPT-4o: Sharp, stabbing pain on movement.

\par\noindent$\bullet$~HyperCLOVA: Pain when changing posture; even at rest there's a
throbbing sensation.

\par\noindent$\bullet$~Llama: Sharp pain in the right flank on movement.

\textbf{Q9.} Was the urine color normal? Any red tint or clots?

\par\noindent$\bullet$~Human: It was fine until yesterday, but this morning the urine came
out a bit red.

\par\noindent$\bullet$~Claude: This morning when I urinated there was a slight red blood
tinge. It's a new symptom and I'm worried.

\par\noindent$\bullet$~GPT-4o: This morning a slight red blood tinge appeared in the urine.

\par\noindent$\bullet$~HyperCLOVA: This morning when I urinated a slight red blood tinge
appeared.

\par\noindent$\bullet$~Llama: No, a slight red blood tinge appeared in the urine.

\textbf{Q10.} Any pain or discomfort when urinating?

\par\noindent$\bullet$~Human: Now that you mention it, I have been urinating frequently
lately.

\par\noindent$\bullet$~Claude: Yes, for the past week I've been urinating frequently with
some tightness and discomfort. I thought it was from the pregnancy.

\par\noindent$\bullet$~GPT-4o: There was some tightness and discomfort when urinating.

\par\noindent$\bullet$~HyperCLOVA: For the past week I've been urinating frequently with
some tightness and discomfort.

\par\noindent$\bullet$~Llama: No, but I had felt some tightness and discomfort when
urinating.

\textbf{Q11.} Any signs of inflammation or bleeding from the vaginal area?

\par\noindent$\bullet$~Human: No, none.

\par\noindent$\bullet$~Claude: No, no inflammation or bleeding from the vaginal area.

\par\noindent$\bullet$~GPT-4o: No, no inflammation or bleeding from the vaginal area.

\par\noindent$\bullet$~HyperCLOVA: No, none.

\par\noindent$\bullet$~Llama: No, no vaginal inflammation or bleeding.

\textbf{Q12.} Are you pregnant?

\par\noindent$\bullet$~Human: Yes, I am 20 weeks pregnant.

\par\noindent$\bullet$~Claude: Yes, I'm 20 weeks pregnant. So far the pregnancy has been
progressing well.

\par\noindent$\bullet$~GPT-4o: Yes, I'm 20 weeks pregnant.

\par\noindent$\bullet$~HyperCLOVA: Yes, 20 weeks pregnant.

\par\noindent$\bullet$~Llama: Yes, I'm 20 weeks pregnant.

\textbf{Q13.} Is this your first pregnancy? Have you had previous
pregnancies?

\par\noindent$\bullet$~Human: Yes, this is my first.

\par\noindent$\bullet$~Claude: This is my first pregnancy. I've never had a miscarriage or
termination.

\par\noindent$\bullet$~GPT-4o: Yes, this is my first pregnancy. I've had no previous
pregnancies, miscarriages, or terminations.

\par\noindent$\bullet$~HyperCLOVA: Yes, this is my first.

\par\noindent$\bullet$~Llama: Yes, first pregnancy. No previous pregnancies.

\textbf{Q14.} Are you on any regular medications?

\par\noindent$\bullet$~Human: No, just Tylenol that I bought yesterday for the fever.

\par\noindent$\bullet$~Claude: No regular medications. I just took Tylenol yesterday for the
fever.

\par\noindent$\bullet$~GPT-4o: No, no regular medications.

\par\noindent$\bullet$~HyperCLOVA: Just Tylenol I bought yesterday; no other regular
medications.

\par\noindent$\bullet$~Llama: No, no regular medications.

\textbf{Q15.} Have you had any major illness in the past, or been diagnosed
with diabetes?

\par\noindent$\bullet$~Human: No, nothing in particular.

\par\noindent$\bullet$~Claude: No history of major illness or diabetes diagnosis. I haven't
had abnormal findings on antenatal exams.

\par\noindent$\bullet$~GPT-4o: No major illness. No diabetes diagnosis.

\par\noindent$\bullet$~HyperCLOVA: No major illness. No diabetes either.

\par\noindent$\bullet$~Llama: No major illness, no diabetes diagnosis.

\textbf{Q16.} Have you ever had cystitis or a urinary-tract infection before?

\par\noindent$\bullet$~Human: No, never.

\par\noindent$\bullet$~Claude: No, no history of cystitis or urinary-tract infection.

\par\noindent$\bullet$~GPT-4o: No.

\par\noindent$\bullet$~HyperCLOVA: No.

\par\noindent$\bullet$~Llama: No, never.

\textbf{Q17.} Have you been drinking less water or eating less recently?

\par\noindent$\bullet$~Human: Yes, since pregnancy I drink frequently, so I've been
deliberately drinking a bit less.

\par\noindent$\bullet$~Claude: I've been drinking a bit less water lately because I was
urinating so often, but my appetite is normal.

\par\noindent$\bullet$~GPT-4o: I've been drinking less water lately because of the frequent
urination.

\par\noindent$\bullet$~HyperCLOVA: Yes, I've been drinking less water because of frequent
urination.

\par\noindent$\bullet$~Llama: Yes, I've been drinking less water lately because of frequent
urination.

\textbf{Q18.} Any recent international travel or contact with someone who had
infectious symptoms?

\par\noindent$\bullet$~Human: No.

\par\noindent$\bullet$~Claude: No international travel or contact with sick people.

\par\noindent$\bullet$~GPT-4o: No.

\par\noindent$\bullet$~HyperCLOVA: No.

\par\noindent$\bullet$~Llama: No, no recent travel or contact with infectious patients.

\textbf{Q19.} Do you drink alcohol or smoke?

\par\noindent$\bullet$~Human: I used to drink occasionally, but quit since pregnancy.

\par\noindent$\bullet$~Claude: I drink only one beer occasionally and never since pregnancy.
I don't smoke.

\par\noindent$\bullet$~GPT-4o: No alcohol or smoking since pregnancy.

\par\noindent$\bullet$~HyperCLOVA: I rarely drank --- just one beer --- and have quit since
pregnancy. I don't smoke.

\par\noindent$\bullet$~Llama: No alcohol since pregnancy. No smoking.

\textbf{Q20.} Have you had sexual relations with your husband recently?

\par\noindent$\bullet$~Human: Yes, I've started carefully a few weeks ago after my
obstetrician said it was safe.

\par\noindent$\bullet$~Claude: Yes, after my obstetrician said it was safe a few weeks ago we
resumed carefully.

\par\noindent$\bullet$~GPT-4o: Yes, we've resumed carefully a few weeks ago after the
obstetrician's advice.

\par\noindent$\bullet$~HyperCLOVA: Yes, we resumed carefully a few weeks ago.

\par\noindent$\bullet$~Llama: Yes, we resumed carefully after the obstetrician advised it was
safe.

\textbf{Q21.} Thank you. Based on the symptoms, your fever is most likely
from acute pyelonephritis. This is a kidney infection, often starting as
cystitis and ascending the ureter. Does that make sense?

\par\noindent$\bullet$~Human: Yes, I understand. So what should I do?

\par\noindent$\bullet$~Claude: Yes, I see. So a urinary-tract infection that ascended to the
kidney has caused the fever and flank pain.

\par\noindent$\bullet$~GPT-4o: Yes, I understand. So a kidney infection causes the fever and
flank pain. Thank you.

\par\noindent$\bullet$~HyperCLOVA: Yes. Cystitis ascending to the kidney caused the symptoms,
then?

\par\noindent$\bullet$~Llama: Yes, I understand.

\textbf{Q22.} For a pregnant patient with acute pyelonephritis like yours,
antibiotic treatment is the principle for both maternal and fetal
safety. You should also drink about 2 L of water daily, not hold urine,
and take antibiotics regularly. Any questions?

\par\noindent$\bullet$~Human: Wait --- is the antibiotic safe for the baby?

\par\noindent$\bullet$~Claude: I'm worried whether antibiotics will be harmful to the baby.
Are pregnancy-safe medications available?

\par\noindent$\bullet$~GPT-4o: Yes, I understand. But I'm worried about the safety of
antibiotics for the baby.

\par\noindent$\bullet$~HyperCLOVA: Will antibiotics affect the baby?

\par\noindent$\bullet$~Llama: Yes. But I'm worried about whether antibiotics may harm the
baby.

\textbf{Q23.} All medications carry some risk, but if pyelonephritis is left
untreated the inflammation can endanger the baby more. We start
treatment early with the safest available antibiotics --- risk varies by
drug, and we'll choose ones with the lowest known risk for fetal
abnormalities and safety concerns.

\par\noindent$\bullet$~Human: Alright. I'll trust your judgement.

\par\noindent$\bullet$~Claude: I understand. So we'll use safer antibiotics whose
pregnancy-related risks have been assessed. Thank you for the
explanation.

\par\noindent$\bullet$~GPT-4o: Yes, I understand. So untreated infection is more dangerous
than antibiotic risk, and you'll use safer options.

\par\noindent$\bullet$~HyperCLOVA: I see. So treating the infection is more important than
not treating it.

\par\noindent$\bullet$~Llama: Yes, I understand.

\subsection*{Case: Chronic cough \textnormal{\textit{(translated from Korean)}}}

\textbf{Q1.} Hello, I'm medical student Cho Su-jin. May I have your name and
age, please?

\par\noindent$\bullet$~Human: Hello, I'm Lee Yun-jung, 55 years old.

\par\noindent$\bullet$~Claude: Hello, doctor. My name is Lee Yun-jung and I'm 55 years old.

\par\noindent$\bullet$~GPT-4o: Hello, I'm Lee Yun-jung, 55 years old.

\par\noindent$\bullet$~HyperCLOVA: My name is Lee Yun-jung, 55 years old.

\par\noindent$\bullet$~Llama: Hello! My name is Lee Yun-jung, 55 years old.

\textbf{Q2.} Hello, Ms. Lee. What brings you in today?

\par\noindent$\bullet$~Human: My cough won't stop. It's been more than three months, so I
came in.

\par\noindent$\bullet$~Claude: I came in because of a cough that's been ongoing for three
months. I thought it was nothing at first, but it gradually got worse.

\par\noindent$\bullet$~GPT-4o: I have a cough that's been continuing for three months. It's
mostly in the morning, and it's a dry cough.

\par\noindent$\bullet$~HyperCLOVA: I came in because of a cough lasting three months. At
first I thought it was a cold, but it's gone on so long that I'm
worried.

\par\noindent$\bullet$~Llama: Lately I've had a cough for three months. It's worse in the
morning and is a dry cough. There's almost no sputum, but sometimes I
feel something stuck in my throat.

\textbf{Q3.} That sounds difficult. Let me ask a few questions to help
diagnose. When during the day do you cough most?

\par\noindent$\bullet$~Human: Mostly in the early morning or just after dawn.

\par\noindent$\bullet$~Claude: I cough mostly in the early morning. It tends to be a bit more
severe when I wake up.

\par\noindent$\bullet$~GPT-4o: Mostly early in the morning.

\par\noindent$\bullet$~HyperCLOVA: Mostly in the morning.

\par\noindent$\bullet$~Llama: Mostly in the morning, especially early. Then around lunchtime,
and a bit less in the evening.

\textbf{Q4.} Are there particular situations where the cough comes --- at
night, after eating, or other?

\par\noindent$\bullet$~Human: When I eat spicy food or overeat, the cough seems worse.

\par\noindent$\bullet$~Claude: Now that I think about it, after meals when my chest feels
heavy and burning the cough is worse. Spicy or oily foods make it worse,
and at night too.

\par\noindent$\bullet$~GPT-4o: After meals when my chest feels heavy and burning the cough
comes. It's worse at night.

\par\noindent$\bullet$~HyperCLOVA: I do cough at night, but more after meals --- especially
with oily or spicy food.

\par\noindent$\bullet$~Llama: After meals, especially after oily or spicy food. Also at
night, when stomach acid feels like it's coming up.

\textbf{Q5.} Do you have sputum or runny nose with the cough?

\par\noindent$\bullet$~Human: No sputum or runny nose.

\par\noindent$\bullet$~Claude: Sputum is rare. It's mostly a dry cough. Sometimes I feel
something stuck in my throat. No runny nose.

\par\noindent$\bullet$~GPT-4o: No sputum or runny nose. Sometimes I feel something stuck in
my throat.

\par\noindent$\bullet$~HyperCLOVA: Sputum is rare; sometimes I feel something stuck in my
throat. No runny nose.

\par\noindent$\bullet$~Llama: No, sputum is rare. No runny nose either. Just the cough.

\textbf{Q6.} Have you had fever or chills?

\par\noindent$\bullet$~Human: No, no fever or cold symptoms --- just the cough.

\par\noindent$\bullet$~Claude: No fever, chills, or body aches with the cough.

\par\noindent$\bullet$~GPT-4o: No fever or chills.

\par\noindent$\bullet$~HyperCLOVA: Nothing like that. Just the cough.

\par\noindent$\bullet$~Llama: No, no fever, chills, or body aches with the cough.

\textbf{Q7.} Any post-nasal drip --- feeling that mucus is dripping down the
back of your throat?

\par\noindent$\bullet$~Human: No, that doesn't happen.

\par\noindent$\bullet$~Claude: Hmm\... I haven't noticed that. No particular sensation of
mucus dripping.

\par\noindent$\bullet$~GPT-4o: No, no such sensation.

\par\noindent$\bullet$~HyperCLOVA: No.

\par\noindent$\bullet$~Llama: No, no post-nasal drip. Sometimes I feel something stuck in my
throat, but only occasionally.

\textbf{Q8.} Any vomiting, diarrhoea, or constipation?

\par\noindent$\bullet$~Human: No.

\par\noindent$\bullet$~Claude: No vomiting, diarrhoea, or constipation. My bowel habits are
unchanged.

\par\noindent$\bullet$~GPT-4o: No, no such symptoms.

\par\noindent$\bullet$~HyperCLOVA: No, but I've had a burning sensation in my chest,
especially after meals.

\par\noindent$\bullet$~Llama: No, no vomiting, diarrhoea, or constipation.

\textbf{Q9.} Does the cough get worse when you lie down?

\par\noindent$\bullet$~Human: It gets worse on waking, so when lying down it tends to come
more.

\par\noindent$\bullet$~Claude: Yes, the cough is worse when lying down. Especially at night
when I try to sleep.

\par\noindent$\bullet$~GPT-4o: Yes, it seems worse when lying down.

\par\noindent$\bullet$~HyperCLOVA: Lying down doesn't particularly make it worse.

\par\noindent$\bullet$~Llama: No, the cough comes when sitting or standing. Lying down too,
but mostly worse at night after oily or spicy food.

\textbf{Q10.} Have you had any swelling or shortness of breath?

\par\noindent$\bullet$~Human: No.

\par\noindent$\bullet$~Claude: No swelling or shortness of breath. I'm living normally with
no shortness of breath on exercise.

\par\noindent$\bullet$~GPT-4o: No swelling or shortness of breath.

\par\noindent$\bullet$~HyperCLOVA: No swelling, no shortness of breath.

\par\noindent$\bullet$~Llama: No, no swelling or shortness of breath. No chest pain.

\textbf{Q11.} What is your height and weight?

\par\noindent$\bullet$~Human: I'm 160 cm and 54 kg.

\par\noindent$\bullet$~Claude: Height 160 cm, weight 54 kg.

\par\noindent$\bullet$~GPT-4o: 160 cm, 54 kg.

\par\noindent$\bullet$~HyperCLOVA: 160 cm, 54 kg.

\par\noindent$\bullet$~Llama: 160 cm, 54 kg.

\textbf{Q12.} Has your weight changed compared to six months ago?

\par\noindent$\bullet$~Human: No, no significant change.

\par\noindent$\bullet$~Claude: No, my weight hasn't really changed over the past six months.

\par\noindent$\bullet$~GPT-4o: No.

\par\noindent$\bullet$~HyperCLOVA: No.

\par\noindent$\bullet$~Llama: No, no weight change.

\textbf{Q13.} Any burning sensation in the chest, or stomach acid coming up?

\par\noindent$\bullet$~Human: Yes, especially after spicy food my chest feels burning.

\par\noindent$\bullet$~Claude: Yes, after meals --- especially spicy ones --- my chest feels
burning, and sometimes I feel acid rising into my throat.

\par\noindent$\bullet$~GPT-4o: Yes, particularly after spicy food my chest feels burning.

\par\noindent$\bullet$~HyperCLOVA: Yes, my chest feels burning after meals, and sometimes I
feel acid rising.

\par\noindent$\bullet$~Llama: Yes, after spicy food my chest feels burning, and acid rises
into my throat.

\textbf{Q14.} Are there triggers like cold air, dust, or allergens that make
the cough worse?

\par\noindent$\bullet$~Human: Not particularly.

\par\noindent$\bullet$~Claude: Not particularly. Cold air, dust, or pollen doesn't seem to
make it worse.

\par\noindent$\bullet$~GPT-4o: No, no specific environmental triggers.

\par\noindent$\bullet$~HyperCLOVA: No.

\par\noindent$\bullet$~Llama: No, no environmental triggers like cold air, dust, or pollen.

\textbf{Q15.} Does the cough come and go with the seasons, with runny nose at
certain times?

\par\noindent$\bullet$~Human: No, it's not seasonal.

\par\noindent$\bullet$~Claude: No, it's not particularly seasonal. No previous diagnosis of
allergic rhinitis.

\par\noindent$\bullet$~GPT-4o: No, no seasonal pattern.

\par\noindent$\bullet$~HyperCLOVA: No, no seasonal pattern.

\par\noindent$\bullet$~Llama: No, no seasonal cough or runny nose.

\textbf{Q16.} Do you have any pre-existing lung conditions, allergic
rhinitis, or GERD diagnosis?

\par\noindent$\bullet$~Human: No, I've never had a lung disease or rhinitis diagnosis.

\par\noindent$\bullet$~Claude: No previous diagnosis of lung disease, allergic rhinitis, or
GERD.

\par\noindent$\bullet$~GPT-4o: No, no such diagnoses.

\par\noindent$\bullet$~HyperCLOVA: No.

\par\noindent$\bullet$~Llama: No, no diagnoses of lung disease, allergic rhinitis, or GERD.

\textbf{Q17.} Are you on any regular medications?

\par\noindent$\bullet$~Human: I take amlodipine for high blood pressure for the past 5 years.

\par\noindent$\bullet$~Claude: Yes, amlodipine for hypertension, for about five years. The
blood pressure is well controlled.

\par\noindent$\bullet$~GPT-4o: Yes, I take amlodipine daily for hypertension.

\par\noindent$\bullet$~HyperCLOVA: Yes, amlodipine for blood pressure.

\par\noindent$\bullet$~Llama: Yes, amlodipine for hypertension.

\textbf{Q18.} Does anyone in your family have lung cancer, or any major lung
or bronchial disease?

\par\noindent$\bullet$~Human: No, no family members with lung disease.

\par\noindent$\bullet$~Claude: No family member has lung cancer or major lung disease. My
mother has the same hypertension and takes medication.

\par\noindent$\bullet$~GPT-4o: No, no family members with lung disease.

\par\noindent$\bullet$~HyperCLOVA: No.

\par\noindent$\bullet$~Llama: No, no family members with lung disease.

\textbf{Q19.} Do you smoke or drink alcohol?

\par\noindent$\bullet$~Human: I've never smoked. Just a little drink at gatherings.

\par\noindent$\bullet$~Claude: I've never smoked. Drink occasionally --- only at company
gatherings.

\par\noindent$\bullet$~GPT-4o: No smoking. Occasional drinking at gatherings.

\par\noindent$\bullet$~HyperCLOVA: No smoking. Drink occasionally.

\par\noindent$\bullet$~Llama: I've never smoked. I drink only at company gatherings.

\textbf{Q20.} What is your occupation? Are there toxic substances or dust in
your work environment?

\par\noindent$\bullet$~Human: I work in office management at a supermarket. I'm mostly in
the office.

\par\noindent$\bullet$~Claude: I work in office administration at a mart. The work
environment is normal --- no toxic substances or dust.

\par\noindent$\bullet$~GPT-4o: I work in office administration at a mart. No toxic substance
or dust exposure.

\par\noindent$\bullet$~HyperCLOVA: I work at a mart in office administration; the environment
isn't dusty.

\par\noindent$\bullet$~Llama: I work in office administration at a mart, mostly in the
office. No toxic or dusty environment.

\textbf{Q21.} Have you had recent contact with someone with severe fever or
cough?

\par\noindent$\bullet$~Human: No, no recent contacts with sick people.

\par\noindent$\bullet$~Claude: No, no recent contact with people who had severe cold or
fever.

\par\noindent$\bullet$~GPT-4o: No.

\par\noindent$\bullet$~HyperCLOVA: No.

\par\noindent$\bullet$~Llama: No, no recent contact with infectious patients.

\textbf{Q22.} Thank you. Based on what you've described, the most likely
cause of your cough is GERD-related cough. Even without pneumonia or
bronchitis, when stomach acid refluxes, it can trigger cough. Does that
make sense?

\par\noindent$\bullet$~Human: Yes, I understand. So GERD is causing the cough?

\par\noindent$\bullet$~Claude: Yes, I understand. So GERD is causing the cough, not lung
disease.

\par\noindent$\bullet$~GPT-4o: Yes, I understand. So acid reflux is the suspected cause of
the cough.

\par\noindent$\bullet$~HyperCLOVA: Yes, GERD-related cough.

\par\noindent$\bullet$~Llama: Yes, I understand. So GERD is causing the chronic cough.

\textbf{Q23.} For diagnosis, an endoscopy to check the oesophagus and
possibly a 24-hour acid-monitoring probe may be needed.

\par\noindent$\bullet$~Human: Alright, I understand.

\par\noindent$\bullet$~Claude: Yes, I understand. Endoscopy and possibly 24-hour acid
monitoring.

\par\noindent$\bullet$~GPT-4o: Yes, I understand.

\par\noindent$\bullet$~HyperCLOVA: Yes, I understand.

\par\noindent$\bullet$~Llama: Yes, I understand. Endoscopy and possibly acid monitoring.

\textbf{Q24.} And since GERD is suspected, for prevention and improvement:
sleep with a slightly elevated pillow, avoid overeating and irregular
meals, and try to avoid coffee, soft drinks, and juices.

\par\noindent$\bullet$~Human: Alright, I'll try.

\par\noindent$\bullet$~Claude: Yes, I'll try the elevated pillow, regular meals, and avoid
coffee and soft drinks.

\par\noindent$\bullet$~GPT-4o: Yes, I'll try those lifestyle changes.

\par\noindent$\bullet$~HyperCLOVA: Yes, I'll do that.

\par\noindent$\bullet$~Llama: Yes, I'll try those lifestyle changes.
\section{Per-Case Encounter-Level Likert Scores}

This appendix provides per-case mean Likert scores with 95\% confidence intervals for the six encounter-level metrics (Comprehensibility, Hallucination-free, Coherence, Consistency, Engagement, Overall Satisfaction), broken down across the three OSCE cases. The pooled-across-cases values are reported in Table~3 of the main paper. Per-case radar plots (Figures D1--D3) visualize the same data to make per-case profile differences across the five sources directly comparable.

\begin{table}[H]
  \centering\scriptsize
  \caption{Per-case mean encounter-level Likert score [95\% bootstrap CI] for each source $\times$ metric, broken down by OSCE case (n = 17 ratings per cell).}
  \label{tab:d1}
  \begin{tabular}{lccccc}
    \toprule
    \textbf{Metric} & \textbf{Human} & \textbf{Claude 3.5} & \textbf{GPT-4o} & \textbf{Llama-3.1} & \textbf{HyperCLOVA} \\
                    & \textbf{(reference)} & \textbf{Sonnet} & & \textbf{70B} & \textbf{X} \\
    \midrule
    \multicolumn{6}{l}{\textit{Hematuria}} \\
    \quad Comprehensibility    & 4.82 [4.65--5.00] & 4.53 [4.24--4.82] & 4.41 [3.94--4.76] & 4.53 [4.24--4.82] & 4.71 [4.47--4.88] \\
    \quad Hallucination-free   & 4.76 [4.53--4.94] & 4.24 [3.76--4.65] & 4.29 [3.94--4.65] & 4.59 [4.35--4.82] & 4.41 [4.12--4.71] \\
    \quad Coherence            & 4.71 [4.41--4.94] & 3.71 [3.35--4.12] & 3.41 [3.00--3.82] & 3.88 [3.41--4.35] & 4.35 [4.00--4.65] \\
    \quad Consistency          & 4.71 [4.47--4.88] & 4.41 [4.00--4.76] & 4.53 [4.24--4.82] & 4.53 [4.24--4.82] & 4.59 [4.29--4.88] \\
    \quad Engagement           & 3.71 [3.12--4.24] & 3.53 [3.12--4.00] & 3.35 [2.94--3.82] & 3.71 [3.23--4.18] & 4.35 [3.94--4.71] \\
    \quad Overall Satisfaction & 4.59 [4.24--4.88] & 3.59 [3.06--4.06] & 3.18 [2.71--3.71] & 3.94 [3.47--4.41] & 4.12 [3.82--4.41] \\
    \midrule
    \multicolumn{6}{l}{\textit{Fever (in pregnancy)}} \\
    \quad Comprehensibility    & 4.88 [4.71--5.00] & 4.65 [4.35--4.88] & 4.76 [4.53--4.94] & 3.18 [2.71--3.65] & 4.41 [4.12--4.71] \\
    \quad Hallucination-free   & 4.65 [4.41--4.88] & 4.59 [4.29--4.88] & 4.59 [4.29--4.88] & 3.71 [3.18--4.18] & 4.12 [3.76--4.47] \\
    \quad Coherence            & 4.94 [4.82--5.00] & 4.41 [4.06--4.71] & 4.41 [4.12--4.71] & 2.59 [2.24--2.94] & 4.06 [3.65--4.41] \\
    \quad Consistency          & 4.88 [4.71--5.00] & 4.76 [4.53--4.94] & 4.65 [4.41--4.88] & 3.53 [3.00--4.12] & 4.41 [4.18--4.65] \\
    \quad Engagement           & 4.24 [3.94--4.53] & 4.00 [3.59--4.35] & 3.76 [3.35--4.18] & 2.53 [2.12--3.00] & 3.88 [3.47--4.29] \\
    \quad Overall Satisfaction & 4.76 [4.47--5.00] & 4.24 [3.71--4.65] & 4.12 [3.82--4.41] & 2.24 [1.82--2.76] & 3.71 [3.29--4.06] \\
    \midrule
    \multicolumn{6}{l}{\textit{Chronic cough}} \\
    \quad Comprehensibility    & 4.82 [4.65--5.00] & 4.18 [3.82--4.53] & 4.59 [4.24--4.88] & 3.41 [3.06--3.82] & 4.59 [4.29--4.88] \\
    \quad Hallucination-free   & 4.06 [3.65--4.41] & 4.24 [3.88--4.59] & 4.35 [3.94--4.71] & 3.53 [3.12--3.94] & 3.53 [3.06--4.00] \\
    \quad Coherence            & 4.76 [4.53--4.94] & 3.88 [3.35--4.35] & 4.41 [4.06--4.71] & 2.35 [2.00--2.71] & 4.29 [3.94--4.65] \\
    \quad Consistency          & 4.65 [4.24--4.94] & 4.00 [3.53--4.47] & 4.41 [4.06--4.76] & 3.41 [2.94--3.88] & 4.18 [3.82--4.53] \\
    \quad Engagement           & 4.59 [4.29--4.82] & 3.94 [3.41--4.47] & 4.12 [3.76--4.47] & 2.59 [2.06--3.12] & 4.24 [3.88--4.59] \\
    \quad Overall Satisfaction & 4.47 [4.12--4.76] & 3.65 [3.12--4.18] & 4.18 [3.76--4.53] & 1.94 [1.59--2.35] & 3.82 [3.35--4.29] \\
    \bottomrule
  \end{tabular}
\end{table}

\begin{figure}[H]
  \centering
  \includegraphics[width=0.55\linewidth]{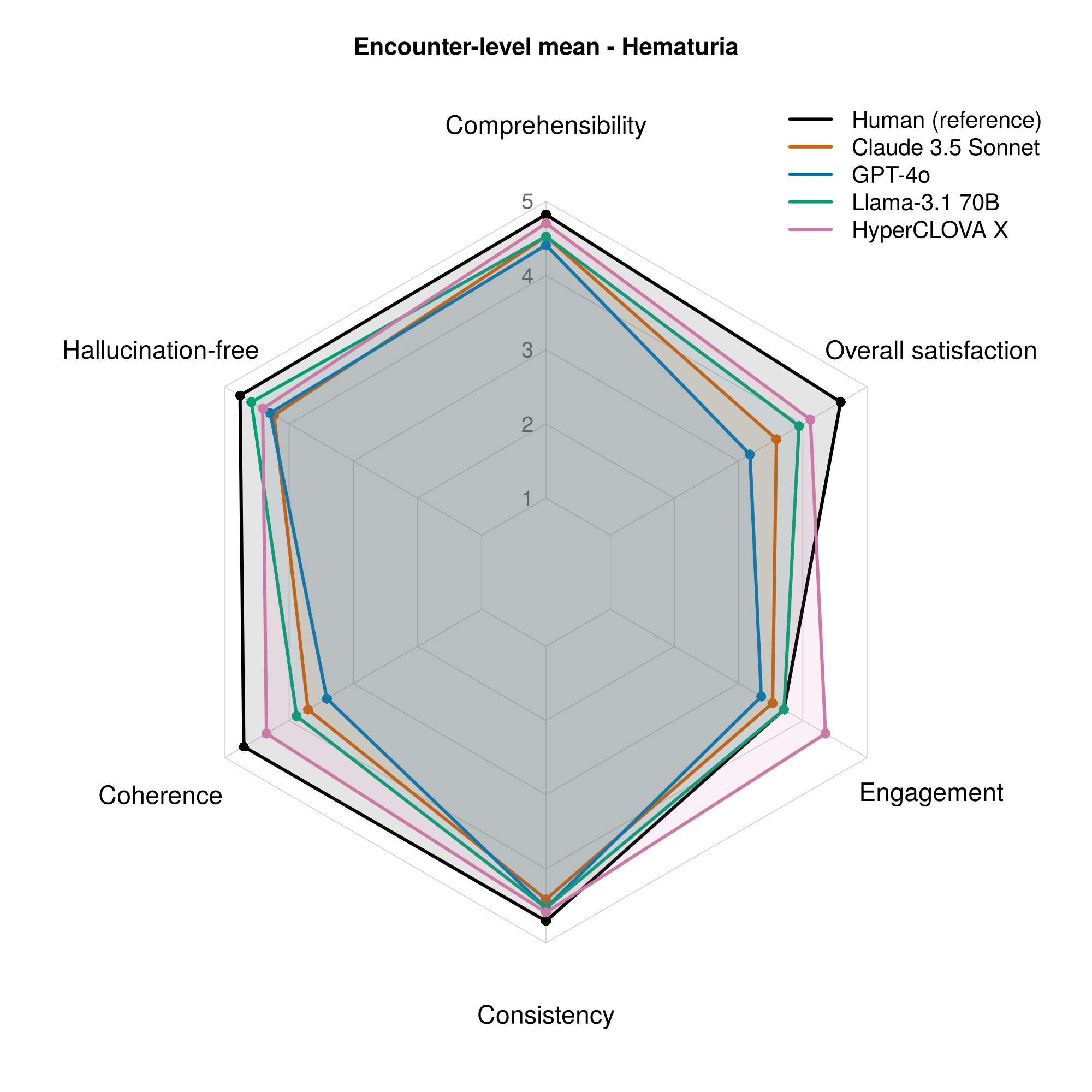}
  \caption{Per-case encounter-level Likert profile --- Hematuria. Each axis is one of the six encounter-level metrics; values plotted are mean Likert scores (n = 17 raters). Five overlaid polygons compare Human (reference), Claude 3.5 Sonnet, GPT-4o, Llama-3.1 70B, and HyperCLOVA X.}
  \label{fig:d1}
\end{figure}

\begin{figure}[H]
  \centering
  \includegraphics[width=0.55\linewidth]{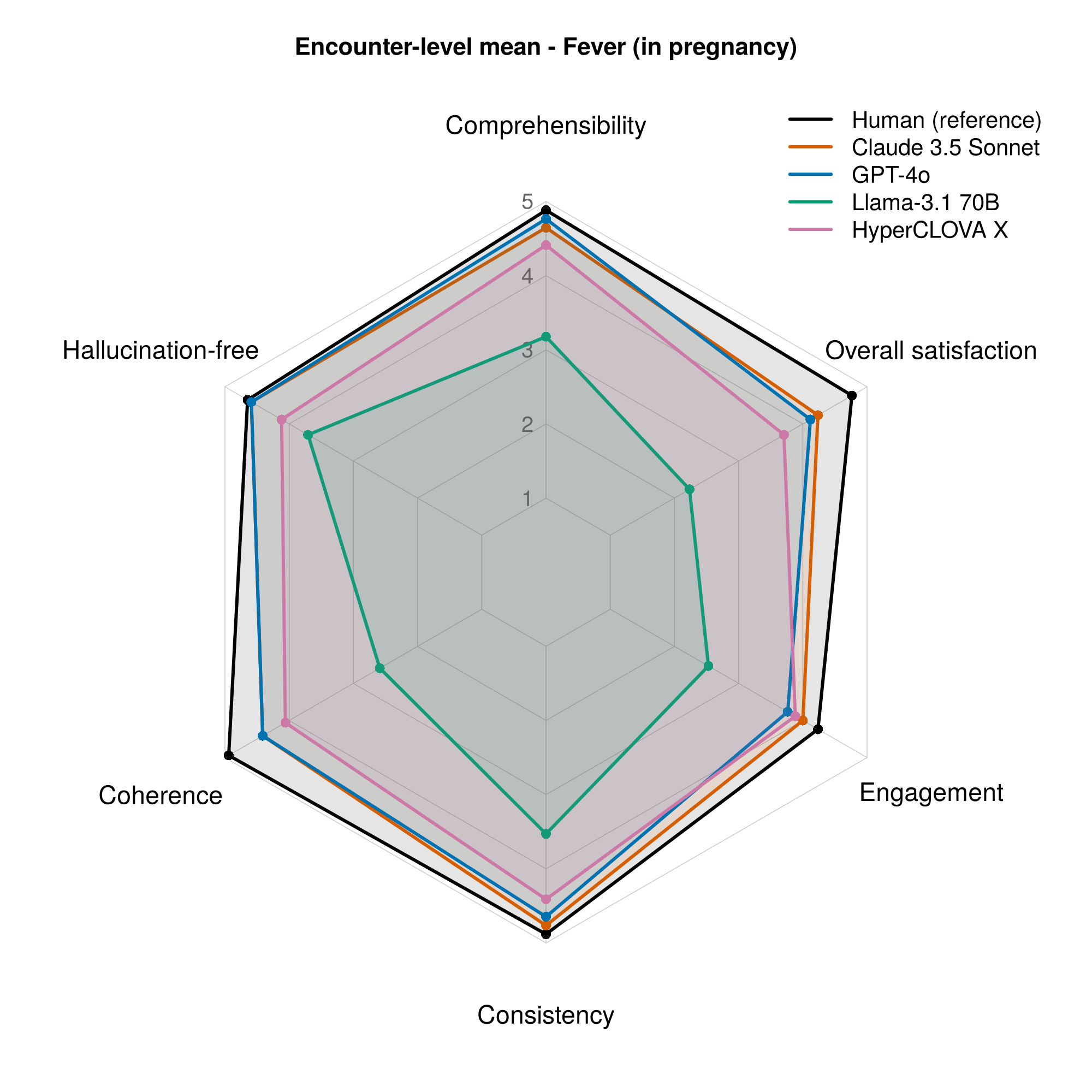}
  \caption{Per-case encounter-level Likert profile --- Fever in pregnancy. Axes and overlay sources are the same as in Figure~\ref{fig:d1}.}
  \label{fig:d2}
\end{figure}

\begin{figure}[H]
  \centering
  \includegraphics[width=0.55\linewidth]{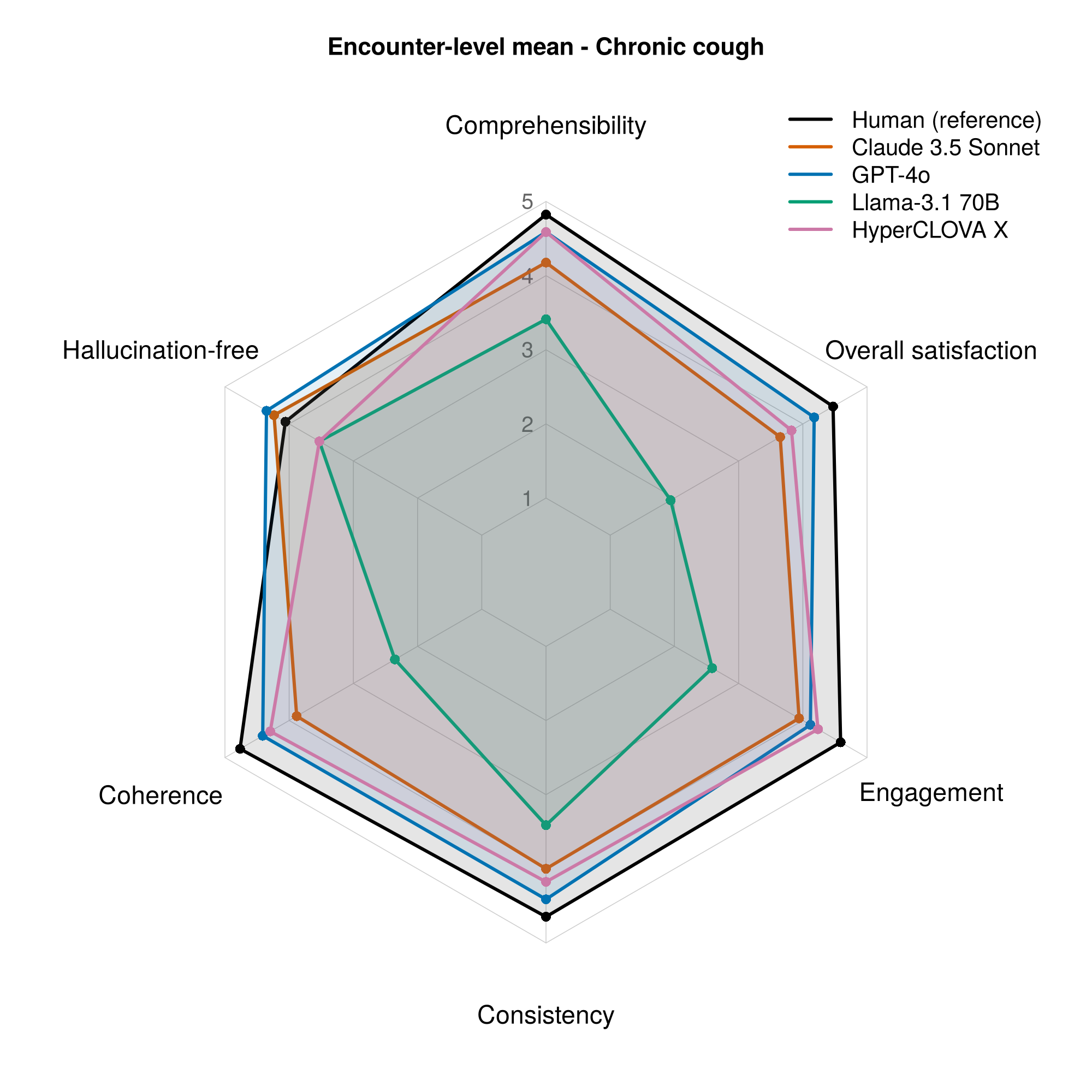}
  \caption{Per-case encounter-level Likert profile --- Chronic cough. Axes and overlay sources are the same as in Figure~\ref{fig:d1}.}
  \label{fig:d3}
\end{figure}

\section{Inter-Rater Reliability and Linguistic-Form Means}

This appendix provides the full inter-rater reliability statistics referenced in \S5.4 and the Source $\times$ Linguistic-form interaction figure referenced in \S5.3.

\begin{table}[H]
  \centering\footnotesize
  \caption{Inter-rater reliability across 17 medical-student raters. ICC(2,k) is reported with 95\% confidence intervals (F-distribution method); Krippendorff's $\alpha$ uses ordinal weights.}
  \label{tab:e1}
  \begin{tabular}{llccc}
    \toprule
    \textbf{Level} & \textbf{Metric / Source} & \textbf{$n$\textsubscript{targets}} & \textbf{ICC(2,k) [95\% CI]} & \textbf{Krippendorff $\alpha$} \\
    \midrule
    Encounter & Comprehensibility      & 15  & 0.91 [0.82--0.96] & 0.26 \\
    Encounter & Hallucination-free     & 15  & 0.79 [0.61--0.91] & 0.14 \\
    Encounter & Coherence              & 15  & 0.94 [0.89--0.98] & 0.41 \\
    Encounter & Consistency            & 15  & 0.83 [0.68--0.93] & 0.15 \\
    Encounter & Engagement             & 15  & 0.86 [0.73--0.94] & 0.22 \\
    Encounter & Overall Satisfaction   & 15  & 0.93 [0.86--0.97] & 0.35 \\
    Encounter & \textbf{Fully pooled}  & 90  & \textbf{0.91 [0.88--0.94]} & \textbf{0.30} \\
    \midrule
    Sentence  & Human (reference)      & 210 & 0.80 [0.75--0.84] & 0.08 \\
    Sentence  & Claude 3.5 Sonnet      & 210 & 0.81 [0.77--0.85] & 0.18 \\
    Sentence  & GPT-4o                 & 210 & 0.82 [0.78--0.86] & 0.15 \\
    Sentence  & Llama-3.1 70B          & 210 & 0.92 [0.90--0.94] & 0.33 \\
    Sentence  & HyperCLOVA X           & 210 & 0.80 [0.76--0.84] & 0.11 \\
    \bottomrule
  \end{tabular}
\end{table}

\begin{figure}[H]
  \centering
  \includegraphics[width=0.85\linewidth]{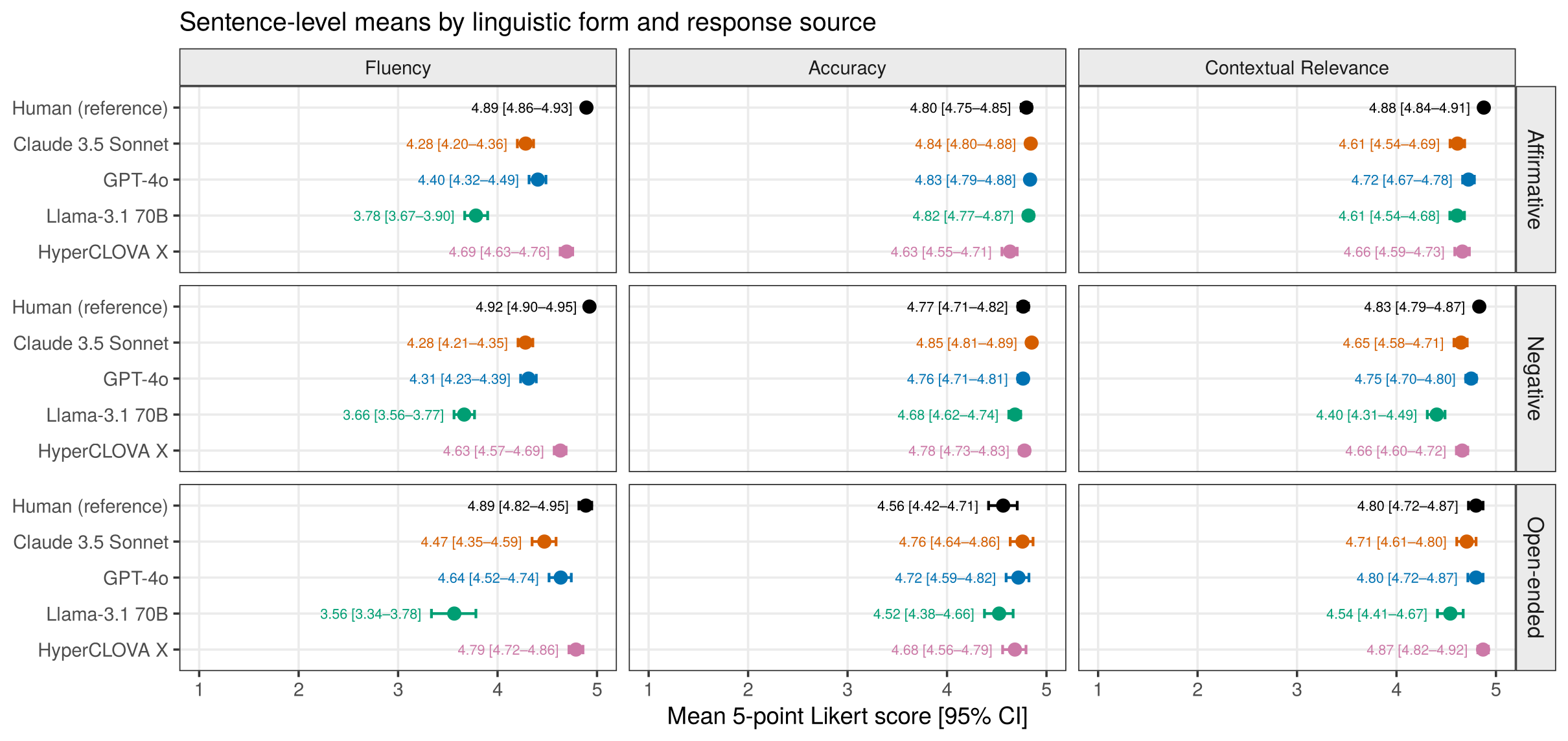}
  \caption{Sentence-level mean Likert score by linguistic form and response source. Panel rows are question forms (Affirmative, Negative-polarity, Open-ended), and panel columns are the three sentence-level metrics; within each panel the y-axis lists the five response sources, and the x-axis is the mean score. Points are means with 95\% bootstrap confidence intervals.}
  \label{fig:e1}
\end{figure}

\section{Platform Corpus Category Distribution}

This appendix reports the per-category distribution referenced in \S6.2.

\begin{table}[H]
  \centering\small
  \caption{Distribution of the eleven history-taking categories in the
  70-prompt evaluation set and in the platform-collected corpus (49
  sessions, 27 scenarios, 1{,}763 utterances). Values are n (\%);
  platform percentages are of the 1{,}614 utterances the scheme covers.
  Non-question directives and Rapport-building talk types fall outside
  the scheme and are reported as counts only. Platform utterances were
  labeled with the same rule-based tagger as the prompt set and
  corrected in full by one researcher.}
  \label{tab:f1}
  \begin{tabular}{lcc}
    \toprule
    \textbf{Category} & \textbf{Prompt set} & \textbf{Platform} \\
    \midrule
    Greeting \& ID check            & 3 (4.3)   & 62 (3.8)   \\
    Chief complaint (C.C.)          & 3 (4.3)   & 29 (1.8)   \\
    HPC --- Character               & 10 (14.3) & 302 (18.7) \\
    HPC --- Associated symptoms     & 21 (30.0) & 463 (28.7) \\
    HPC --- Aggravating/relieving   & 3 (4.3)   & 106 (6.6)  \\
    Past medical history            & 6 (8.6)   & 200 (12.4) \\
    Drug history                    & 3 (4.3)   & 75 (4.6)   \\
    Family history                  & 2 (2.9)   & 68 (4.2)   \\
    Social history                  & 9 (12.9)  & 157 (9.7)  \\
    Clinical reasoning              & 2 (2.9)   & 58 (3.6)   \\
    Patient education               & 8 (11.4)  & 94 (5.8)   \\
    \midrule
    \textbf{Total within the scheme} & \textbf{70} & \textbf{1{,}614} \\
    \midrule
    \multicolumn{3}{l}{\textit{Outside the scheme}} \\
    Non-question directives          & ---       & 67         \\
    Rapport-building talk           & ---       & 82         \\
    \bottomrule
  \end{tabular}
\end{table}

\end{document}